\documentclass[12pt, draftclsnofoot, onecolumn]{IEEEtran}
\usepackage{graphicx,amssymb,amsmath}
\usepackage[noadjust]{cite}%cite
\usepackage{setspace}               % set line space
\usepackage{stfloats}

\usepackage{amsthm}
\usepackage{amsmath}
\usepackage{flushend}
\usepackage{cases,subeqnarray}
\usepackage{bm,multirow,bigstrut}
\usepackage{textcomp}
\usepackage{latexsym,bm}
\usepackage{booktabs,changebar}
\usepackage{xcolor}
\usepackage{mathtools}
\usepackage{dsfont}
\usepackage{extarrows}
\usepackage{mathrsfs}
\usepackage{cite}
\usepackage{bm}
\usepackage{cleveref}
\usepackage{multicol}       % table
\usepackage{multirow}       % table
\usepackage{array}          % table
\usepackage{colortbl}
\usepackage{makecell}
\usepackage{xcolor}
\definecolor{crimson}{RGB}{192,0,0}         % color crimson
\definecolor{navy}{RGB}{47,85,151}         % color crimson

\makeatletter                               % algorithm
\newif\if@restonecol
\makeatother

\makeatletter
\newif\if@restonecol
\makeatother

\usepackage[linesnumbered,ruled,vlined]{algorithm2e}
\usepackage{algpseudocode}
\usepackage{amsmath}
\renewcommand{\arraystretch}{1.5} %

\theoremstyle{plain}

\newtheorem{lemm}{Lemma}

\theoremstyle{plain}

\newcommand{\argmin}[1]{{\underset{{#1}}{\mathrm{arg\,min}}}}
\newcommand{\vect}[1]{\mathbf{#1}}

\def\diag{\mathrm{diag}}

\def\Htran{\mbox{\tiny $\mathrm{H}$}}
\def\Ttran{\mbox{\tiny $\mathrm{T}$}}
\def\CN{\mathcal{N}_{\mathbb{C}}} %Complex Gaussian
\begin{document}

%----------------------------title&author&thanks----------------------------
\title{Energy-Efficient Cell-Free Massive MIMO Through Sparse Large-Scale Fading Processing}
\author{
%\vspace{-1.0em}\setlength{\baselineskip}{11pt}
Shuaifei~Chen,~\IEEEmembership{Graduate Student Member,~IEEE}, Jiayi~Zhang,~\IEEEmembership{Senior Member,~IEEE}, Emil Bj{\"o}rnson,~\IEEEmembership{Fellow,~IEEE}, {\"O}zlem Tu{\u{g}}fe Demir,~\IEEEmembership{Member,~IEEE}, and Bo~Ai,~\IEEEmembership{Fellow,~IEEE}\\
%\vspace{-3em}
\thanks{This article was presented in part at the IEEE SPAWC 2022 \cite{chen2022sparse}. This work was supported in part by National Key R\&D Program of China under Grant 2020YFB1807201, in part by National Natural Science Foundation of China under Grants 61971027 and 62221001, in part by Beijing Natural Science Foundation under Grant L202013, in part by Natural Science Foundation of Jiangsu Province, Major Project under Grant BK20212002. E.~Bj\"ornson was supported by the FFL18-0277 grant from the Swedish Foundation for Strategic Research.

Shuaifei Chen is with the School of Electronic and Information Engineering and the Frontiers Science Center for Smart Highspeed Railway System, Beijing Jiaotong University, Beijing 100044, China, and also with Purple Mountain Laboratories, Nanjing 211111, China (e-mail: shuaifeichen@bjtu.edu.cn).

Jiayi Zhang is with the School of Electronic and Information Engineering and the Frontiers Science Center for Smart Highspeed Railway System, Beijing Jiaotong University, Beijing 100044, China. (e-mail: jiayizhang@bjtu.edu.cn).

Emil Bj{\"o}rnson is with the Department of Computer Science, KTH Royal Institute of Technology, 164 40 Kista, Sweden (e-mail: emilbjo@kth.se).

{\"O}zlem Tu{\u{g}}fe Demir was with the Department of Computer Science, KTH Royal Institute of Technology, 164 40 Kista, Sweden. She is now with the Department of Electrical and Electronics Engineering, TOBB University of Economics and Technology, Ankara, Turkey  (e-mail: ozlemtugfedemir@etu.edu.tr).

Bo Ai is with the State Key Laboratory of Rail Traffic Control and Safety, Beijing Jiaotong University, Beijing 100044, China, and also with the Henan Joint International Research Laboratory of Intelligent Networking and Data Analysis, Zhengzhou University, Zhengzhou 450001, China (e-mail: boai@bjtu.edu.cn).}
}

\maketitle
%\thispagestyle{empty}   % no page number for the first page
%----------------------------abstract----------------------------
\begin{abstract}
  Cell-free massive multiple-input multiple-output (CF mMIMO) systems serve the user equipments (UEs) by geographically distributed access points (APs) by means of joint transmission and reception. To limit the power consumption due to fronthaul signaling and processing, each UE should only be served by a subset of the APs, but it is hard to identify that subset. Previous works have tackled this combinatorial problem heuristically.
  In this paper, we propose a sparse distributed processing design for CF mMIMO, where the AP-UE association and long-term signal processing coefficients are jointly optimized.
  We formulate two sparsity-inducing mean-squared error (MSE) minimization problems and solve them by using efficient proximal approaches with block-coordinate descent.
  For the downlink, more specifically, we develop a virtually optimized large-scale fading precoding (V-LSFP) scheme using uplink-downlink duality.
  The numerical results show that the proposed sparse processing schemes work well in both uplink and downlink. In particular, they achieve almost the same spectral efficiency as if all APs would serve all UEs, while the energy efficiency is 2-4 times higher thanks to the reduced processing and signaling.
%\vspace{-1.5em}
\end{abstract}

%----------------------------keywords----------------------------
%\vspace{-1.2em}

%\newpage

\section{Introduction}%\vspace{-0.2em}

As the number of active wireless devices is steadily growing \cite{index2019cisco}, the increasing requirements and demands for wireless communications force academia and industry to consider not only ``how much and fast" the information can be transferred but also ``how green" the networks can become in terms of the energy efficiency (EE).
This shift in perception makes EE as important as a performance metric as spectral efficiency (SE) for fifth-generation (5G) networks \cite{han2020energy}. During data transmission, the EE is defined as the ratio between the data rate and total power consumption \cite{bjornson2017massive}.
Cellular massive multiple-input multiple-output (mMIMO) with access points (APs) equipped with large antenna arrays became the key technology for simultaneously improving the SE and EE in 5G \cite{marzetta2010noncooperative,larsson2014massive,andrews2014will,parkvall2017nr}.
Looking towards the future, the main limiting factors for the SE and EE have now become the inter-cell interference caused by lack of cooperation between the APs, the large pathlosses between the APs and the user equipments (UEs) when using a small number of elevated APs, and the internal hardware energy consumption of the APs themselves \cite{survey2016buzzi}.
The sixth-generation (6G) networks are expected to improve the SE and EE gains by $100 \times$ over 5G networks \cite{saad2019vision} and must address these issues.
This requires a denser network infrastructure operating in a cell-free (CF) manner that shifts the network from cell-centric to user-centric, and thus, provides ubiquitous coverage, improved network SE, and improved EE \cite{zhang2019multiple,chen2021survey,demir2022cell}.

In the past few years, user-centric CF mMIMO has attracted extensive attention from the research community \cite{cellfreebook}.
This paradigm inherits the {\it interference suppression gain} enabled by multiple antennas per AP from Cellular mMIMO and improves the {\it macro-diversity gain} by increasing the AP deployment density.
In CF mMIMO systems, a large number of distributed APs are collaborating through a central processing unit (CPU) to serve the UEs with coherent joint transmission and reception, as illustrated in Fig.~\ref{fig:system}. This increases the average and worst-case data rates and reduces the total power consumption.
Thus, CF mMIMO is envisioned as a promising paradigm shift for 6G networks \cite{saad2019vision}. The key difference from previous coordinated multipoint approaches is the dense deployment, user-centric approach, and signal processing schemes inherited from Cellular mMIMO.

Due to the UE-AP-CPU architecture, the signal processing tasks in CF mMIMO systems can be distributed between the APs and the CPU in different ways \cite{nayebi2016performance,bjornson2019making}.
According to how many of the tasks are delegated to the APs, CF mMIMO can operate in a {\it centralized} or {\it distributed} manner.
In the centralized operation, the APs act as relays between the UEs and the CPU, which performs channel estimation and all signal processing by exploiting the instantaneous channel state information (CSI) gathered from the APs via the fronthaul connections.
Although the centralized operation exhibits higher {\it user-experienced} data rates (i.e., $95\%$-likely SE) than its distributed alternative, it requires much higher computational complexity.
As illustrated in Fig.~\ref{fig:system}, the alternative distributed operation is a two-stage processing procedure, in which {\it each} AP locally performs channel estimation and  signal processing based on those estimates, while the CPU is only responsible for the final or initial processing of data using scaling factors that only depend on the large-scale fading (LSF) coefficients.
This two-stage technique of processing was originally proposed for Cellular mMIMO, where the central unit linearly combines messages from/to each AP corresponding to the UEs from different cells to effectively eliminate inter-cell interference.
This is referred to as {\it LSF decoding (LSFD)} \cite{adhikary2017uplink} for the uplink and as {\it LSF precoding (LSFP)} \cite{ashikhmin2018interference,demir2020large} for the downlink.
For CF mMIMO, early papers on the topic of distributed uplink operation proposed to simply take the average of the local messages from different APs at the CPU.
This can perform poorly since it neglects the inter-AP LSF information that is also available at the CPU and some APs can do more harm than good when serving far-away UEs.
With this consideration, the authors in \cite{nayebi2016performance} developed LSFD for uplink CF mMIMO.
When it comes to the downlink, CF mMIMO {\it inherently} performs LSFP since the transmitted messages for different APs are all encoded at the CPU but scaled differently by the APs when doing power allocation.
Therefore, the concept of the LSFP was not mentioned in the existing CF mMIMO literature.
To demonstrate the connections, the terminology ``LSFP"  is anyway used to represent the two-stage downlink transmit power allocation.

Although the distributed operation of CF mMIMO achieves a good compromise between data rates and computational complexity compared to the fully centralized operation \cite{cellfreebook}, it might not be energy efficient in its original form where {\it all APs serve all UEs} \cite{nayebi2016performance,bjornson2019making}.
It is unnecessary for an AP to waste its power, computational, and fronthaul resources to serve distant UEs (with weak channels) when those UEs have better channels to other APs \cite{buzzi2019user}.
The geometry induces a sparse structure on the practically meaningful AP-UE associations.
Prior works have suggested associating each UE with a subset of APs in advance and then excluding APs not associated with this UE when computing the LSFD vector \cite{bjornson2020scalable,chen2020structured,demir2021cell}.
Since the problem is combinatorial, to our best of knowledge, only heuristic methods have been proposed; see \cite{cellfreebook} for a recent survey.
However, treating the AP-UE association as a separate combinatorial problem from the LSFD design, which is employed to maximize the SE \cite{nayebi2016performance}, is suboptimal.
This motivates us to consider the association as a part of the uplink LSFD and downlink LSFP design and employ of sparsity-inducing methods to jointly solve the association problem and signal processing design.

Sparse optimization methods have many successful applications in the fields of signal processing, image processing, and computer vision \cite{zhang2015survey}.
Specific to wireless communications, sparse optimization has been applied for random access \cite{liu2018sparse}, activity detection \cite{chen2018sparse}, and node sleeping \cite{van2020joint}.
For example, the authors in \cite{van2020joint} shut down some ``unnecessary" APs in a CF mMIMO system while satisfying the requested SEs by formulating the sparse reconstruction problem as a mixed-integer second-order cone program, where the globally optimal solution is found by utilizing the branch-and-bound approach.
Similarly, in \cite{demir2022cell}, mixed-binary programming is exploited to activate only the minimal subsets of APs for each UE to reduce the end-to-end network power consumption, where CF mMIMO is implemented on top of a virtualized cloud radio access network.

\subsection{Main Contributions}

We develop an energy-efficient distributed processing framework for CF mMIMO systems, which makes use of sparsity methods but in a novel way. We formulate a new sparse optimization problem for CF mMIMO that minimizes the data mean-squared-error (MSE), to enforce sparsity on the LSFD and LSFP coefficients.
In consequence, the data rates are barely deteriorated while the power consumption needed to achieve it is minimized, which leads to higher EE.
Our major contributions are listed as follows:
\begin{itemize}

  \item We propose the sparse LSF processing design for both uplink and downlink, where joint AP-UE association and LSFD/LSFP is achieved by formulating sparsity-inducing MSE-minimizing problems to push small LSFD/LSFP coefficients to zero.
      We consider two kinds of sparsity: element-wise (EW) and group-wise (GW).
  \item We solve these formulated sparsity problems efficiently by developing proximal algorithms with block-coordinate descent (BCD).
      The proposed algorithms contain closed-form updates and, thus, operate faster than the well-used optimization tool CVX \cite{cvx2015}.
  \item We develop a novel virtually optimized LSFP (V-LSFP) scheme for downlink power allocation in CF mMIMO systems by using the uplink-downlink duality. It is interesting to achieve $1.7\times$ $95\%$-likely downlink SE compared to the benchmark using distributed fractional power allocation (FPA) \cite{interdonato2019scalability,bjornson2020scalable}.
  \item We compare the proposed sparse schemes with their fully-connected alternatives (where all APs serve all UEs) \cite{nayebi2016performance}, and their partial alternatives \cite{cellfreebook,chen2020structured} with the separate AP-UE association as in \cite{bjornson2020scalable}. % considering both SE and EE, different combining/precoding schemes, and different AP deployment setups.
      The simulation results show that the proposed sparse LSF schemes significantly improve the EE with only a slight SE loss compared to the benchmarks.

\end{itemize}

The conference version of this paper, \cite{chen2022sparse}, only considers the sparse LSFD (S-LSFD) design in the uplink with GW sparsity. Herein, we extend \cite{chen2022sparse} to a more generalized case considering both uplink and downlink with EW and GW sparsities.

\subsection{Paper Outline and Notation}

The remainder of this paper is organized as follows.
Section \ref{sec:system} introduces the system model for our considered CF mMIMO system.
Section \ref{sec:uplink} elaborates on the distributed uplink transmissions with LSFD.
The sparse processing with sparse optimization is developed in Section \ref{sec:problem} by formulating two sparsity-inducing problems.
Section \ref{sec:downlink} extends the analysis and design to the downlink where the LSFP and corresponding sparse processing are proposed. In Section~\ref{sec:power}, the details of the power consumption model are provided along with the definition of energy efficiency.
Section \ref{sec:results} numerically evaluates the proposed schemes and compares them with the considered benchmarks.
Finally, we draw the conclusions and implications in Section \ref{sec:conclusion}.

\subsubsection{Reproducible Research}
All the simulation results can be reproduced using the Matlab code and data files available at: https://github.com/ShuaifeiChen273/sparse-LSFprocess-CFmMIMO.

\subsubsection{Notation}

Boldface lowercase letters, $\bf x$, denote column vectors, boldface uppercase letters, $\bf X$, denote matrices, and calligraphic uppercase letters, $\cal A$, denote sets.
${\bf I }_n$ denotes the $n \!\times \! n$ identity matrix.
The superscripts $^{\Ttran}$, $^\star$, and $^{\Htran}$ denote the transpose, conjugate, and conjugate transpose, respectively.
$x_i \!=\! [{\bf x}]_i$, $(x)_+ \!=\! \max(x,0)$, and ${\rm{sign}}(\cdot)$ is the signum function.
${\mathbb E}\{ \cdot \}$ computes the expected values and ${\cal N}_{\mathbb C}\left({{\bf 0},{\bf R}}\right)$ denotes the multi-variate circularly symmetric complex Gaussian distribution with correlation matrix $\bf R$.

\section{CF mMIMO System Model}\label{sec:system}

\begin{figure}[t!]
%\vspace{0.5em}
\centering
\includegraphics[scale=0.95]{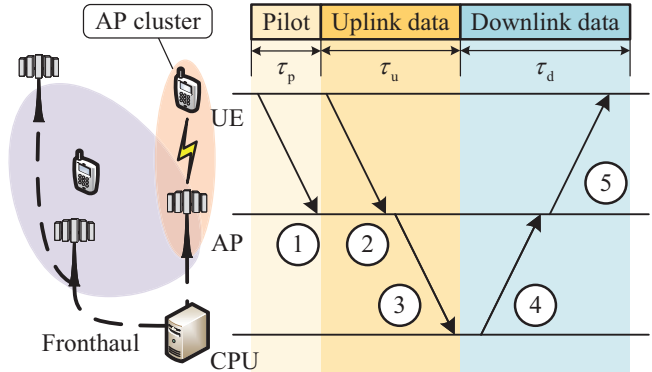}
\caption{Each UE is served by a subset of the APs in our considered user-centric CF mMIMO system. With the distributed operation, the signal processing tasks are divided between the APs and the CPU as indicated for 1) channel estimation, 2) local receive combining, 3) data decoding, 4) data encoding, and 5) local transmit precoding.
\label{fig:system}}
\end{figure}

We consider a CF mMIMO system that consists of $K$ single-antenna UEs and $L$ geographically distributed APs, each equipped with $N$ antennas.
We adopt the user-centric CF architecture, where each UE is served by a subset of the APs, as illustrated in Fig.~\ref{fig:system}.
The AP subsets of different UEs may overlap and are selected based on the UEs' channel qualities and service requirements. We will optimize these subsets but for now, we denote by ${\cal D}_l \subset \{{ 1,\ldots, K }\}$ the subset of UEs served by AP $l$ and denote by ${\cal M}_k \subset \{{ 1,\ldots,L }\}$ the subset of APs serving UE $k$.
All APs are connected via fronthaul connections to a CPU, which are coordinating the signal processing of all UEs, while the actual processing is distributed over the APs.

We adopt the standard time division duplex (TDD) operation and block fading model, where the time-frequency resources are divided into coherence blocks so that the channel coefficients can be assumed fixed in each block. We consider spatially correlated Rayleigh fading, which implies that the channel between AP $l$ and UE $k$ denoted by ${\bf h}_{kl} \in {\mathbb C}^N$ takes an independent realization in each coherence block according to
\begin{equation}
  {{\bf{h}}_{kl}} \sim {\cal N}_{\mathbb C} ({\bf 0}, {\bf R}_{kl}),
\end{equation}
where ${\bf{R}}_{kl} \in {\mathbb C}^{N \times N}$ is the spatial correlation matrix and $\beta_{kl} \buildrel \Delta \over = {\rm tr}({\bf R}_{kl})/N$ is the LSF coefficient describing pathloss and shadowing.
It is assumed that AP $l$ knows the correlation matrices $\{{\bf R}_{kl} : \forall k\}$ of all UEs since these represent the long-term channel statistics \cite{bjornson2017massive}.

Each coherence block is used for both uplink and downlink payload data transmission and some portion is also used for uplink pilots.
More precisely, each coherence block of $\tau_{\rm c}$ channel uses is divided into three phases: a) $\tau_{\rm p}$ channel uses are dedicated for pilot transmission and channel estimation; b) $\tau_{\rm u}$ channel uses for uplink payload data; and c) the remaining $\tau_{\rm d} = \tau_{\rm c} - \tau_{\rm p} - \tau_{\rm u}$ channel uses for downlink payload data.
We adopt the two-stage distributed processing approach in this paper \cite{cellfreebook} (as illustrated in Fig.~\ref{fig:system}), where only the data decoding and encoding are delegated to the CPU.
The other signal processing tasks are done at the APs.

\subsection{Uplink Pilot Transmission and Channel Estimation}

During the channel estimation, each AP locally estimates the channels based on the uplink pilot transmission from the UEs.
We consider a mutually orthogonal set of $\tau_{\rm p}$ pilot sequences that must be shared between the UEs because in practical large networks, we will likely have  $\tau_{\rm p} \ll K$.
We denote by $t_k$ the index of the pilot assigned to UE $k$ and by ${\cal S}_{t_k}$ the set of UEs sharing pilot $t_k$.
When the UEs in ${\cal S}_{t_k}$ transmit pilot $t_k$, the received signal ${\bf{y}}_{{t_k}l}^{\rm{p}} \in {{\mathbb C}^N}$ at AP $l$ (after taking the inner product of the received signal and the pilot sequence $t_k$) is \cite[Sec. 3]{bjornson2017massive}
\begin{equation}\label{eq:received pilot signal at AP}
  {\bf y}_{{t_k}l}^{\rm{p}} = \sum_{i \in {{\cal S}_{t_k}}} {\sqrt {{\tau _{\rm p}}{p _{\rm{p}}}} {{\bf{h}}_{il}}}  + {{\bf{n}}_{{t_k}l}},
\end{equation}
where ${\bf n}_{t_k l} \sim {\cal N}_{\mathbb C}( {\bf 0}, \sigma^2 {\bf I}_N )$ is the receiver noise with noise power $\sigma^2$ and $p_{\rm{p}}$ is the pilot transmit power of each UE.
The \emph{minimum MSE (MMSE)} estimate of ${\bf h}_{kl}$ is \cite[Sec. 3]{bjornson2017massive}
\begin{equation}\label{eq:mmse estimate}
  {\widehat{\bf h}}_{kl} = \sqrt{\tau_{\rm p} p_{\rm p}} {\bf R}_{kl} {\bf \Psi}_{t_k l}^{-1} {\bf y}_{t_k l}^{\rm p} \sim {\cal N}_{\mathbb C}( {\bf 0}, {\bf B}_{kl} ),
\end{equation}
where ${\bf \Psi}_{t_k l} = {\mathbb E}\{ {\bf y}_{t_k l}^{\rm p} ({\bf y}_{t_k l}^{\rm p})^{\Htran} \} = \sum_{i \in {\cal S}_{t_k}} \tau_{\rm p} p_{\rm p} {\bf R}_{il} + \sigma^2 {\bf I}_N$
is the correlation matrix of ${\bf y}_{t_k l}^{\rm p}$ in \eqref{eq:received pilot signal at AP} and ${\bf B}_{kl} = \tau_{\rm p} p_{\rm p} {\bf R}_{kl} {\bf \Psi}_{t_k l}^{-1} {\bf R}_{kl}$.

\section{Uplink Data Transmissions with LSFD}\label{sec:uplink}

In this section, we provide the details of the distributed implementation of uplink reception, which are needed to formulate our design problem.
Each AP locally employs an arbitrary receive combining scheme to obtain local soft estimates of the UE data.
These estimates are then gathered at the CPU, which combines them using the LSFD approach.

In the uplink data phase, the received signal ${\bf y}_{l}^{\rm{ul}} \in \mathbb{C}^N$ at AP $l$ is a superposition of the signals  from all UEs:
\begin{equation}
	{\bf y}_{l}^{\rm{ul}} = \sum_{i=1}^{K} {\bf h}_{il}s_{i} + {\bf n}_{l},
\end{equation}
where $s_{i} \in \mathbb{C}$ is the signal transmitted by UE $i$, $p_i = \mathbb{E} \{ |s_i|^2 \}$ is the corresponding transmit power, and ${\bf n}_{l}  \sim \CN({\bf 0},\sigma^2{\bf I}_{N})$ is the independent additive receiver noise.
AP $l$ selects the normalized local combining vector ${\bf v}_{kl} =  {\bar{\bf v}}_{kl}/ \sqrt{{\mathbb E}\{ \| {\bar{\bf v}}_{kl} \|_2^2 \}} \in \mathbb{C}^N$ for UE $k$ and then computes its local estimate of $s_k$ as
\begin{equation}
\widehat s_{kl} = \vect{v}_{kl}^{\Htran}  {\bf y}_{l}^{\rm{ul}}.
\end{equation}
One good option is to use the {\it local MMSE (L-MMSE)} combining scheme \cite{bjornson2020scalable}
\begin{equation} \label{eq:lmmse combining}
\bar{\bf v}_{kl} = p_k \left( \sum_{i=1}^K p_i \left({\widehat{\bf h}}_{il} {\widehat{\bf h}}_{il}^{\Htran} + {\bf R}_{il} - {\bf B}_{il}\right) + \sigma^2 {\bf I}_N \right)^{-1}{\widehat{\bf h}}_{kl}
\end{equation}
that suppresses interference and minimizes the local MSE ${\mathbb E}\{ |s_k - {\widehat s}_{kl}|^2| \{{\widehat{\bf h}_{il}}: \forall i\}  \}$.
Alternatively, the maximum ratio (MR) processing scheme with $\bar{\bf v}_{kl} = {\widehat{\bf h}}_{kl}$ can be used.
Note that, for a generic UE $k$, although ${\bf v}_{kl} \ne {\bf 0}$ for all APs, only the serving APs in ${\cal M}_k$ need to compute ${\bf v}_{kl}$.

Next, the APs transfer their local data estimates to the CPU, which performs the final decoding of $s_k$ by linearly combining the local estimates:
\begin{equation} \label{eq:global data etimate}
\widehat{s}_k = \sum_{l=1}^{L} a_{kl}^{\star} \widehat{s}_{kl} = \sum_{l=1}^{L} a_{kl}^{\star} \vect{v}_{kl}^{\Htran}  {\bf y}_{l}^{\rm{ul}},
\end{equation}
where $a_{kl} \in \mathbb{C}$ is the weight that the CPU assigns to the local signal estimate $\widehat{s}_{kl}$.
In LSFD, the CPU selects the weights  $\{ a_{kl} \}$ as a deterministic function of the channel statistics (to avoid sharing channel estimates \cite{cellfreebook}).
Note that only those APs assigning a non-zero value to $a_{kl}$ participate in the decoding, thus this formulation supports a user-centric architecture. For a given set $\{a_{kl}\}$  of LSFD weights, the serving APs of UE $k$ can be extracted as ${\cal M}_k = \{l:a_{kl} \ne 0\}$.

By letting
$
{\bf g}_{ki} = [{\bf v}_{k1}^{\Htran}{\bf h}_{i1},  \ldots, {\bf v}_{kL}^{\Htran}{\bf h}_{iL}]^{\Ttran} \in {\mathbb C}^L
$
denote the receive-combined channels from UE $i$ when receiving signals from UE $k$, and
$
{\bf a}_k = [a_{k1},\ldots,a_{kL}]^{\Ttran} \in {\mathbb C}^L
$
denote the LSFD weight vector of UE $k$, the estimate of $s_k$ in \eqref{eq:global data etimate} can be rewritten as
\begin{equation}\label{eq:s etimate}
\widehat{s}_k = {\bf a}_{k}^{\Htran}{\bf g}_{kk}s_{k} + \sum_{i=1, i\ne k}^{K} {\bf a}_{k}^{\Htran}{\bf g}_{ki}s_{i} + n'_k,
\end{equation}
where $n^\prime_{k} = \sum_{l=1}^{L} a_{kl}^{\star} \vect{v}_{kl}^{\Htran}\vect{n}_{l}$ is the resulting noise.
The effective uplink channel $\vect{a}_{k}^{\Htran}\vect{g}_{kk}$ in \eqref{eq:s etimate} is not known at the CPU but its average $\mathbb{E}\{\vect{a}_{k}^{\Htran}  \vect{g}_{kk}\} = \vect{a}_{k}^{\Htran} \mathbb{E}\{ \vect{g}_{kk}\}$ is deterministic and non-zero if the receive combiner is selected as suggested above.
Therefore, it can be assumed to be available at the CPU and we can therefore quantify the achievable uplink SE using the \emph{hardening bound} \cite[Thm. 4.4]{bjornson2017massive}. More precisely, the resulting SE of UE $k$ is
\begin{equation}\label{eq:se}
{\sf SE}^{\rm ul}_k = \frac{\tau_{\rm u}}{\tau_{\rm c}} {\log _2}\left( {1 + {\sf SINR}^{\rm ul}_k} \right) \quad \textrm{bit/s/Hz},
\end{equation}
 where
\begin{equation} \label{eq:uplink sinr}
{\sf SINR}^{\rm ul}_k = \frac{|{\bf a}_k^{\Htran} {\pmb \xi}_k |^2}{{\bf a}_k^{\Htran} {\bf \Delta}_k {\bf a}_k -  |{\bf a}_k^{\Htran} {\pmb \xi}_k |^2} = \frac{|{\bf a}_k^{\Htran} {\pmb \xi}_k |^2} { {\bf a}_k^{\Htran} ( {\bf \Delta}_k - {\pmb \xi}_k {\pmb \xi}_k^{\Htran} ) {\bf a}_k },
\end{equation}
is the effective uplink signal-to-interference-plus-noise ratio (SINR) \cite[Thm. 5.4]{cellfreebook} with
\begin{align}
 {\bf \Delta}_k  & \!=\!  \sum_{i=1}^K p_i {\mathbb E} \{{\bf g}_{ki} {\bf g}_{ki}^{\Htran}\} \!+\! \sigma^2{\bf I}_L \!\in\! {\mathbb C}^{L \times L}, \label{eq:Delta} \\
{\pmb \xi}_k  & \!=\!  \sqrt{p_k} \mathbb{E}\{{\bf g}_{kk}\} \!\in\! {\mathbb C}^{L}. \label{eq:xi}
\end{align}
We note that the effective uplink SINR in \eqref{eq:uplink sinr} is a generalized Rayleigh quotient with respect to ${\bf a}_k$.
Hence, with the help of the generalized eigenvector result \cite[Lem.~B.10]{bjornson2017massive} and matrix inversion lemma \cite[Lem.~B.4]{bjornson2017massive}, the optimal LSFD (O-LSFD) weight vector is
\begin{equation} \label{eq:O-LSFD}
{\bf a}_k^{\rm opt} = c_k {\bf \Delta}_k^{-1} {\pmb \xi}_k
\end{equation}
with $c_k \in {\mathbb C}$ being an arbitrary non-zero scaling factor. The resulting  maximum SINR value is
$
{\sf SINR}_k^{\rm ul} = {\pmb \xi}_k^{\Htran} ({\bf \Delta}_k - {\pmb \xi}_k {\pmb \xi}_k^{\Htran} )^{-1} {\pmb \xi}_k
$.

Moreover, we notice that the uplink MSE in the data decoding of UE $k$ is
\begin{equation}\label{eq:uplink mse}
    {\sf MSE}_k^{\rm ul} = \mathbb{E}\{ |s_k-\widehat{s}_k |^2 \} = {\bf a}_k^{\Htran} {\bf \Delta}_k {\bf a}_k - 2 \sqrt{p_k} \Re ({\bf a}_k^{\Htran} {\pmb \xi}_k)+p_k,
\end{equation}
which is minimized by the LSFD vector
\begin{equation}\label{eq:LSFD MSE}
{\bf a}_k^{\rm mse} = \sqrt{p_k}{\bf \Delta}_k^{-1}{\pmb \xi}_k,
\end{equation}
which is equal to ${\bf a}_k^{\rm opt}$ in \eqref{eq:O-LSFD} if the scaling factor is set to $c_k=\sqrt{p_k}$.
While there is only one LSFD vector minimizing the MSE, we can use any scaling factor to maximizing the SINR.
We conclude that we can identify an optimal LSFD vector by minimizing the MSE instead of maximizing the SINR, which is a feature that we will exploit in the remainder of this paper.

By using the notation ${\bf a } = {[ {{\bf a}_1^{\Ttran}, \ldots ,{\bf a}_K^{\Ttran}} ]^{\Ttran}}\in \mathbb{C}^{KL}$, ${\pmb \xi} = {[ \sqrt{p_1}{\pmb \xi}_1^{\Ttran}, \ldots ,\sqrt{p_K} {\pmb \xi}_K^{\Ttran} ]^{\Ttran}}\in \mathbb{C}^{KL}$, and ${\bf \Delta} = {\diag}( {\bf \Delta}_1, \ldots ,{\bf \Delta}_K )\in \mathbb{C}^{KL\times KL}$, we can express the uplink sum MSE of all UEs as
\begin{equation}\label{eq:uplink sum mse}
    \sum_{k = 1}^K {\sf MSE}_k^{\rm ul} = {\bf a}^{\Htran} {\bf \Delta}{\bf a} - 2\Re ( {\bf a}^{\Htran}{\pmb \xi} ) + \sum_{k = 1}^K {p}_k.
\end{equation}
Recall that in \eqref{eq:uplink mse}, each uplink MSE only depends on the respective UE's LSFD vector ${\bf a}_k$.
Hence, finding the collective LSFD vector ${\bf a}^{\rm opt}$ that minimizes the sum MSE $\sum_{k = 1}^K {\sf MSE}_k^{\rm ul}$
is equivalent to finding the set of O-LSFD vectors $\{{\bf a}_k^{\rm opt}: k = 1,\ldots,K\}$ that simultaneously minimize their corresponding uplink MSEs.

\section{Sparse LSFD with MSE Minimization}\label{sec:problem}

One way to implicitly obtain the AP selection for UE $k$ is to first design a suitable LSFD vector as if all APs serve the UE and then let only the APs with non-zero weights serve it: ${\cal M}_k = \{l:a_{kl} \ne 0,\ l = 1,\ldots,L\}$.
The problem with this approach is that the O-LSFD vector ${\bf a}^{\rm opt}$ in \eqref{eq:O-LSFD} in general only contains non-zero values, so all APs would have to serve all UEs. However, we have noticed that ${\bf a}^{\rm opt}$ typically contains a few large values and many small values due to the natural pathloss differences between APs and UEs in a distributed deployment. In this section, we will propose the S-LSFD design that resembles O-LSFD but pushes small weights to zero, thereby greatly limiting how many APs must serve each UE.

\subsection{Problem Formulation}

We recall that O-LSFD is obtained by minimizing the quadratic form in \eqref{eq:uplink sum mse} with respect to the collective LSFD vector $\bf a$. Inspired by this fact and sparse reconstruction methods, we propose the generic real-valued
MSE minimization problem
\begin{equation}\label{eq:p0}
 \min_{{\underline{\bf a}} \in {\mathbb R}^{2KL}}
 {\underline{\bf a}}^{\Ttran} {\underline{\bf \Delta}}{\underline{\bf a}} - 2  {\underline{\bf a}}^{\Ttran}{\underline{\pmb \xi}}  + \Omega ( {\underline{\bf a}} )
\end{equation}
with the real variables ${\underline{\bf a}} = {[ { {\underline{\bf a}}_1^{\Ttran}, \ldots , {\underline{\bf a}}_K^{\Ttran}} ]^{\Ttran}}\in \mathbb{R}^{2KL}$, ${\underline{\pmb \xi}} = {[ \sqrt{p_1}{\underline{\pmb \xi}}_1^{\Ttran}, \ldots ,\sqrt{p_K} {\underline{\pmb \xi}}_K^{\Ttran} ]^{\Ttran}}\in \mathbb{R}^{2KL}$, and ${\underline{\bf \Delta}} = {\diag}( {\underline{\bf \Delta}}_1, \ldots ,{\underline{\bf \Delta}}_K )\in \mathbb{R}^{2KL\times 2KL}$, where
\begin{equation}\label{eq:comp2real}
\begin{aligned}
 {\underline{\bf a}}_k &=
 \begin{bmatrix}
  \Re( {\bf a}_k )\\
  \Im( {\bf a}_k )
\end{bmatrix}\in {\mathbb R}^{2L},\
 {\underline{\pmb \xi}}_k =
 \begin{bmatrix}
  \Re( {\pmb \xi}_k )\\
  \Im( {\pmb \xi}_k )
\end{bmatrix}\in {\mathbb R}^{2L}, \\
 {\underline{\bf \Delta}}_k &=
 \begin{bmatrix}
  \Re( {\bf \Delta}_k ) & -\Im( {\bf \Delta}_k )\\
  \Im( {\bf \Delta}_k ) & \Re( {\bf \Delta}_k )
\end{bmatrix}\in {\mathbb R}^{2L \times 2L}.
\end{aligned}
\end{equation}
The first two terms in \eqref{eq:p0} ${\underline{\bf a}}^{\Ttran} {\underline{\bf \Delta}}{\underline{\bf a}} - 2  {\underline{\bf a}}^{\Ttran}{\underline{\pmb \xi}} $ represent the ``MSE" cost, which is a convex function of ${\underline{\bf a}}$.
The third term $\Omega ( {\underline{\bf a}} )$ is a sparsity-inducing function that can be designed to encourage small values in ${\underline{\bf a}}$ to become zero at the optimal solution.
We refer to the minimizer of \eqref{eq:p0} as a S-LSFD vector.
By selecting different $\Omega ( {\underline{\bf a}} )$, different sparsity patterns can be achieved in the LSFD vector ${\underline{\bf a}}$.
We will consider two key examples in this section.

\subsection{EW Sparsity and Proximal Algorithm}\label{subsubsec:ew problem}

Element-wise (EW) sparsity  can be induced on the LSFD vector by using the $\ell_1$-norm, as
\begin{equation}  \label{eq:EW-sparsity}
\Omega({\underline{\bf a}}) = \lambda \| {\underline{\bf a}} \|_1,
\end{equation}
where $\lambda \geq 0$ is a tunable EW regularization parameter. %, which limits the number of non-zero values in vector ${\bf a}$.
The physical interpretation behind EW sparsity is to limit the average number of UEs that each AP serves, but otherwise letting the optimization problem select freely which AP-UE associations that should remain.
A large value of $\lambda$ induces more EW sparsity.
When using \eqref{eq:EW-sparsity}, the optimization problem in \eqref{eq:p0} becomes
\begin{equation}\label{eq:p1}
 {\sf P}^{\rm ew}: \quad
 \min_{{\underline{\bf a}} \in {\mathbb R}^{2KL}}
 {\underline{\bf a}}^{\Ttran} {\underline{\bf \Delta}}{\underline{\bf a}} - 2  {\underline{\bf a}}^{\Ttran}{\underline{\pmb \xi}}  + \lambda \| {\underline{\bf a}} \|_1,
\end{equation}
which is convex since the $\ell_1$-norm penalty is a convex function.
In fact, we can solve the $K$ subproblems
\begin{equation}\label{eq:p1.1}
 {\sf P}^{\rm ew}_k: \quad
 \min_{ {\underline{\bf a}}_k \in {\mathbb R}^{2L} }
  f( {\underline{\bf a}}_k )  + \lambda \| {\underline{\bf a}}_k \|_1,\ k=1,\dots,K
\end{equation}
in parallel to obtain the solution to \eqref{eq:p1} since both the ``MSE" cost and sparsity function can be decoupled between the UEs, where $f( {\underline{\bf a}}_k ) = {\underline{\bf a}}_k^{\Ttran} {\underline{\bf \Delta}}_k {\underline{\bf a}}_k - 2\sqrt{p_k}  {\underline{\bf a}}_k^{\Ttran}{\underline{\pmb \xi}}_k$.

Since the subproblems $\{{\sf P}^{\rm ew}_k\}$ in \eqref{eq:p1.1} are convex with non-smooth sparsity-inducing penalties, the proximal methods can be utilized to solve them efficiently \cite{rish2014sparse}.

By using the proximal methods, we start with an initial point ${\underline{\bf a}}_k^{0}$ which can be initialized by its corresponding O-LSFD vector ${\bf a}_k^{\rm opt}$ using \eqref{eq:comp2real}, and then compute a sequence of updates ${\underline{\bf a}}_k^{n}$ that converges to the optimal solution to \eqref{eq:p1.1}, where $n$ is the iteration index.
Given the  ${\underline{\bf a}}_k^{n}$ obtained at iteration $n$, we can find the next update ${\underline{\bf a}}_k^{n+1}$ by solving the following {\it proximal problem}
\begin{equation}\label{eq:p1.3}
\min_{ {\underline{\bf a}}_k \in {\mathbb R}^{2L} }
 \frac{1}{2} \left\| {\underline{\bf a}}_k - G({\underline{\bf a}}_k^{n}) \right\|_2^2 + \mu \lambda \| {\underline{\bf a}}_k \|_1 ,
\end{equation}
where $G({\underline{\bf a}}_k^{n}) = {\underline{\bf a}}_k^{n} - \mu \nabla f( {\underline{\bf a}}_k^{n} )$ is the so-called {\it gradient update} and $\mu$ is the step length which can be computed in practice via line search \cite{rish2014sparse}.
The unique solution of \eqref{eq:p1.3} can be found due to the strong convexity \cite{rish2014sparse}. This is given as follows.
\begin{lemm}
Since $\nabla f( {\underline{\bf a}}_k ) = 2 {\underline{\bf \Delta}}_k {\underline{\bf a}}_k - 2 \sqrt{p_k} {\underline{\pmb \xi}}_k$, the unique solution of \eqref{eq:p1.3} can be obtained as
\begin{equation}
{\rm Prox}_{\mu \lambda,\ell_1} ( G( {\underline{\pmb \alpha}}_k^{n}) ) = \argmin{{\underline{\bf a}}_k \in {\mathbb R}^{2L}} \ \frac{1}{2} \left\| {\underline{\bf a}}_k - G({\underline{\bf a}}_k^{n}) \right\|_2^2 + \mu \lambda \| {\underline{\bf a}}_k \|_1 ,
\end{equation}
which is the proximal operator of the $\ell_1$-norm \cite{rish2014sparse} and can be componentwisely computed  as
\begin{equation}\label{eq:prox l1}
[ {\rm Prox}_{\mu,\ell_1} ({\bf u}) ]_i = {\rm{sign}}( u_i ) \cdot { ( {| u_i | - \mu } )_+ }.
\end{equation}
\end{lemm}
\begin{IEEEproof}
The proof follows the results in \cite{rish2014sparse} and is omitted due to limited space.
\end{IEEEproof}
To obtain the minimizer of ${{\sf P}}^{\rm ew}_k$ in \eqref{eq:p1.1}, the vector can be updated as
\begin{equation}
{\underline{\bf a}}_k^{n+1} \leftarrow {\rm Prox}_{\mu \lambda,\ell_1} ( G( {\underline{\hat{\bf a}}}_k^{n}) )
\end{equation}
with the Nesterov step ${\underline{\hat{\bf a}}}_k^{n} = {\underline{\bf a}}_k^{n} + \frac{n-1}{n+2} ({\underline{\bf a}}_k^{n} - {\underline{\bf a}}_k^{n-1})$ that is known to accelerate the convergence to the solution of \eqref{eq:p1.1}  \cite{rish2014sparse}.
By performing the inverse transformations in \eqref{eq:comp2real}, we achieve the complex-valued EW S-LSFD vectors.

\subsection{GW Sparsity and Proximal Algorithm with BCD}

The EW sparsity approach limits the average number of UEs served by an AP, but without inducing any preference on how the UE load is distributed among the APs.
In practice, we might prefer that some APs are not serving any UEs at all, so that we can save power by putting them into sleep mode.
This property can be encouraged by also inducing group-wise (GW) sparsity on the LSFD vector. More precisely, we propose to use the composite $\ell_1 + \ell_1/\ell_2$-norm to simultaneously induce GW and EW sparsity:
\begin{equation}\label{eq:penalty gw}
\Omega({\underline{\bf a}}) = \gamma \sum_{l=1}^L \| {\pmb x}_{l}\|_2 + \lambda \| {\underline{\bf a}} \|_1, \ {\pmb x}_l = [\Re( {\pmb a}_l )^{\Ttran},  \Im( {\pmb a}_l )^{\Ttran}]^{\Ttran}\in {\mathbb R}^{2K},
\end{equation}
where $\gamma$ is the tunable GW regularization parameter and
${\pmb a}_l = [a_{1l},\ldots,a_{Kl}]^{\Ttran} \in {\mathbb C}^K$ is subset of the vector ${\bf a}$ related to AP $l$.
Larger value of $\gamma$ induces more GW sparsity on vector ${\underline{\bf a}}$.
The first term in \eqref{eq:penalty gw} is a $\ell_1/\ell_2$-norm that behaves like a $\ell_1$-norm applied to the vector $[\| {\pmb x}_{1}\|_2,\ldots,\| {\pmb x}_{L}\|_2 ]^{\Ttran}$. Element $l$, i.e., $\| {\pmb x}_{l}\|_2$, is small if AP $l$ has little impact on the decoding and thus the $\ell_1/\ell_2$-norm promotes making such values identically zero (i.e., inactivate the AP).
The second term limits the number of UEs served by the remaining active APs.
The sparse problem in \eqref{eq:p0} becomes
\begin{equation}\label{eq:p2}
 {\sf P}^{\rm gw}: \quad
 \min_{{\underline{\bf a}} \in {\mathbb R}^{2KL}}
 {\underline{\bf a}}^{\Ttran} {\underline{\bf \Delta}}{\underline{\bf a}} - 2  {\underline{\bf a}}^{\Ttran}{\underline{\pmb \xi}} + \gamma \sum_{l=1}^L \| {\pmb x}_{l}\|_2 + \lambda \| {\underline{\bf a}} \|_1
\end{equation}
which is convex since the composite $\ell_1 + \ell_1/\ell_2$-norm penalty is a convex function.

The $\ell_1/\ell_2$-norm term restricts \eqref{eq:p2} from being decomposed into $K$ subproblems that can be solved in parallel, in contrast to \eqref{eq:p1} in the EW case.
But fortunately, \eqref{eq:p2} is separable between the APs so that the BCD approach can be used to guarantee convergence to the global optimum of \eqref{eq:p2} \cite{rish2014sparse}.
We equivalently rewrite the original problem ${\sf P}^{\rm gw}$ as
\begin{align}\label{eq:p2.1}
 \min_{{\underline{\bf a}} \in {\mathbb R}^{2KL}}
 \left\| {\bar{\pmb \xi}} - \sum_{l=1}^L {\bf X}_l {\pmb x}_l \right\|_2^2 + \gamma \sum_{l=1}^L \| {\pmb x}_{l}\|_2 + \lambda \| {\underline{\bf a}} \|_1,
\end{align}
where ${\bf X}^{\Ttran} {\bf X} = {\underline{\bf \Delta}}$ and ${\bf X}_l \in {\mathbb R}^{2KL \times 2K}$ is the submatrix of ${\bf X}$ with columns corresponding to group $l$ such that ${\bf X}{\underline{\bf a}}=\sum_{l=1}^L{\bf X}_l{\pmb x}_l$.
We use the notation ${\bar{\pmb \xi}} = ({\bf X}^{\Ttran})^{-1} {\underline{\pmb \xi}} \in {\mathbb R}^{2KL}$.
Note that \eqref{eq:p2.1} has the same form as the so-called ``sparse-group Lasso" problem \cite{simon2013sparse}.
Hence, by using the BCD approach, we can solve \eqref{eq:p2.1} efficiently by iteratively minimizing the subproblem of group $l$ while fixing the coefficients of the other groups:
\begin{equation}\label{eq:p2.2}
 {\sf P}^{\rm gw}_l: \quad
 \min_{ {\pmb x}_l \in {\mathbb R}^{2K} }
  g( {{\pmb x}}_l ) + \Omega'({\pmb x}_l),\quad l=1,\dots,L,
\end{equation}
where $g( {{\pmb x}}_l ) = \| {\bf r}_l - {\bf X}_l {\pmb x}_l \|_2^2$, $\Omega'({\pmb x}_l) =  \gamma \| {\pmb x}_{l}\|_2 + \lambda \| {\pmb x}_l \|_1$, and ${\bf r}_l = {\bar{\pmb \xi}} - \sum_{j\ne l} {\bf X}_j {\pmb x}_j$ is the partial residual of ${\bar{\pmb \xi}}$ subtracting all group coefficients except group $l$.
$\Omega'({\pmb x}_l)$ implies that for group $l$, the other group coefficients are considered fixed and their penalties can be ignored.

\begin{algorithm}[t!]
\label{algo:warm_restart}
%\doublespacing
\caption{Algorithm for Warm-Restart}
\KwIn{$\lambda$, $\bar \lambda$, $\eta \in (0,1)$}

\KwOut{${\bf a} \in {\mathbb C}^{KL}$}

$\lambda' = {\bar \lambda}$; \\

(Outer loop) \While{$\lambda' \ge \lambda$}
{
(Inner loop) Solve the considered sparsity problem with $\lambda'$ and update ${\bf a}$;\\

\If{$\lambda' = \lambda$}
{
{\bf{Break}};\\
}
\Else
{
$\lambda' \leftarrow \max(\eta \lambda' , \lambda)$;\\
}
}
\end{algorithm}

Similar to solving ${{\sf P}}^{\rm ew}_k$, for a subproblem ${{\sf P}}^{\rm gw}_l$ of group $l$, given the current ${{\pmb x}}_l^{n}$ obtained at iteration $n$, the next update ${{\pmb x}}_l^{n+1}$ is found by solving the following proximal problem
\begin{equation}\label{eq:p2.4}
\min_{ {{\pmb x}}_l \in {\mathbb R}^{2K} }
 \frac{1}{2} \left\| {{\pmb x}}_l - G( {{\pmb x}}_l^{n}) \right\|_2^2 + \mu \Omega'({{\pmb x}}_l) ,
\end{equation}
where $G( {{\pmb x}}_l^{n}) = {{{\pmb x}}_l^{n} - \mu \nabla {g} ({{\pmb x}}_l^{n} )}$.
The unique solution to \eqref{eq:p2.4} can be found due to the strong convexity \cite{rish2014sparse} and is given as follows.
\begin{lemm}\label{lemm:prox}
Since $\nabla g( {{\pmb x}}_l ) = 2{{\bf X}}_l^{\Ttran}( {{\bf X}}_l {{\pmb x}}_l -  {{\bf r}}_l )$, the unique solution of \eqref{eq:p2.4} can be computed as
\begin{equation}\label{eq:prox_l1+l2}
{\rm Prox}_{\mu,\Omega'} ( G( {{\pmb x}}_l^{n}) ) = {\rm Prox}_{\mu \gamma,\ell_2} \circ {\rm Prox}_{\mu \lambda,\ell_1} ( G( {{\pmb x}}_l^{n}) ),
\end{equation}
where $f \circ g (x) \triangleq f(g(x))$ for any function $f$ and $g$,
\begin{equation}\label{eq:prox l2}
{\rm Prox}_{\mu,\ell_2} ({\bf u}) = \begin{cases}
{\frac{\bf u}{\|{\bf u}\|_2}{ ( \|{\bf u}\|_2 - \mu } )_+,}&{{\rm if}\ {\bf u} \ne {\bf 0},}\\
{\bf 0},&{{\rm otherwise},}
\end{cases}
\end{equation}
is the proximal operator of the $\ell_2$-norm \cite{rish2014sparse}, and ${\rm Prox}_{\mu,\ell_1} ({\bf x})$ is the proximal operator of the $\ell_1$-norm given in \eqref{eq:prox l1}.
\end{lemm}
\begin{IEEEproof}
The proof follows a similar approach as in \cite{simon2013sparse}, but for problem \eqref{eq:p2.1}.
The details are given in Appendix \ref{appe:prox} for completeness.
\end{IEEEproof}\vspace{0.75em}
The minimizer of group $l$ can be updated as
\begin{equation}
{{\pmb x}}_l^{n+1} \leftarrow {\rm Prox}_{\mu,\Omega{'}} ( G( {\hat{{\pmb x}}}_l^{n} ) )
\end{equation}
with the Nesterov step ${\hat{{\pmb x}}}_l^{n} = {{\pmb x}}_l^{n} + \frac{n-1}{n+2} ({{\pmb x}}_l^{n} - {{\pmb x}}_l^{n-1})$ accelerating the convergence \cite{rish2014sparse}, and is then fixed while the other groups are minimized until next iteration.
By iteratively updating $\{ {{\sf P}}^{\rm gw}_l: l=1,\ldots,L \}$, the global solution to \eqref{eq:p2.2} can be reached.
With the inverse transformations in \eqref{eq:penalty gw}, we achieve the complex-valued GW S-LSFD vectors.
Although the mixed sparse penalty in \eqref{eq:penalty gw} is more generalized than the penalty in \eqref{eq:EW-sparsity}, it requires more computational complexity.

\subsection{Algorithm Implementation}\label{subsec:}

\begin{algorithm}[t!]
\label{algo:ew}
%\doublespacing
\caption{Algorithm for solving ${\sf P}^{\rm ew}$}
\KwIn{${\underline{\bf \Delta}}\in {\mathbb R}^{2KL \times 2KL}$, ${\underline{\bf a}} \in {\mathbb R}^{2KL}$, ${\underline{\pmb \xi}} \in {\mathbb R}^{2KL}$, $\lambda$, $\mu$, ${n}_{\max}$}

\KwOut{${{\bf a}} \in {\mathbb C}^{KL}$}

\For{$k = 1,\ldots,K$}
{
$n = 1$; \\
${\underline{\bf a}}_k^{-} \leftarrow {\underline{\bf a}}_k$; \\

\Repeat{$n = n_{\max}$ or convergence}
{
${\underline{\hat{\bf a}}}_k = {\underline{\bf a}}_k + \frac{n-1}{n+2} ({\underline{\bf a}}_k - {\underline{\bf a}}_k^{-})$;\\
Compute ${\rm Prox}_{\mu \lambda,\ell_1} ( G( {\underline{\hat{\bf a}}}_k) )$ with the help of \eqref{eq:prox l1};\\
${\underline{\bf a}}_k \leftarrow {\rm Prox}_{\mu \lambda,\ell_1} ( G( {\underline{\hat{\bf a}}}_k) )$;\\
${\underline{\bf a}}_k^{-} \leftarrow {\underline{\bf a}}_k$;\\
$n \leftarrow n + 1$;\\
}
}
Obtain ${\bf a} \in {\mathbb C}^{KL}$ with the inverse transformation in \eqref{eq:comp2real}.\\
\end{algorithm}

We have noticed that the EW and  GW sparsity problems are faster to solve with a larger regularization parameter $\lambda$.
With this consideration in mind, we propose to perform the warm-restart strategy \cite{rish2014sparse} on $\lambda$, which accelerates the convergence by solving a sequence of simple subproblems.
The warm-restart strategy starts with a large regularization term ${\bar \lambda} \gg \lambda$, then iteratively shrinks ${\bar \lambda}$ towards $\lambda$ and solves the corresponding subproblems.
In each iteration, the subproblem is solved by employing the solution to the previous subproblem as the initialization.
In other words, the warm-restart strategy used in our scenario operates as a sequence of nested loops, which is summarized in Algorithm \ref{algo:warm_restart}.
The considered sparse problems ${\sf P}^{\rm ew}$ and ${\sf P}^{\rm gw}$ can be solved by performing Algorithm \ref{algo:ew} and Algorithm \ref{algo:gw}, respectively.
These algorithms can be initialized by the O-LSFD vectors without sparsity and terminated when the maximum number of iterations $n_{\rm max}$ is reached or convergence, measured by the change in objective function value.

\section{Downlink Transmissions with LSFP}\label{sec:downlink}

In this section, we consider the distributed downlink transmission with the goal of limiting the number of APs that serve each UE and the number of active APs.
The downlink payload data of each UE is first sent to the APs that serve it.
Next, the data symbols are locally precoded at the APs with local precoding vectors designed based on instantaneous channel estimates and then transmitted using AP-specific power coefficients. These coefficients are designed based on long-term statistics and, thus, correspond to LSFP in the Cellular literature \cite{ashikhmin2018interference}.
We extend the sparse optimization to the downlink and develop a sparse LSFP (S-LSFP) design where the joint AP-UE association and LSFP is achieved.

In the downlink data phase, the distributed implementation is realized by constructing a linearly combined precoded signals from each AP.
Let $\varsigma_i \in {\mathbb C}$ denote the unit-power downlink data signal intended for UE $i$ with ${\mathbb E}\{ |\varsigma_i|^2 \} = 1$.
The data signals for different UEs $\{ \varsigma_i: i=1,\ldots,K \}$ are independent.
For a generic AP $l$, the CPU encodes the related symbols $\{ \varsigma_i: i \in {\cal D}_l \}$ and transfers them to AP $l$ via the fronthaul links.
Then, AP $l$ constructs the transmitted signal as
\begin{equation}
	{\bf x}_{l} = \sum_{i=1}^{K} \sqrt{\rho_{il}}{\bf w}_{il} \varsigma_{i},
\end{equation}
where ${\bf w}_{il} =  {\bar{\bf w}}_{il}/ \sqrt{{\mathbb E}\{ \| {\bar{\bf w}}_{il} \|_2^2 \}} \in \mathbb{C}^N$ is the normalized precoding vector that AP $l$ selects for UE $i$ such that ${\mathbb E}\{\|{\bf{w}}_{il}\|_2^2\} =  1$.
The precoding vector ${\bar{\bf w}}_{il}$ can have an arbitrary norm, while ${\bf w}_{il}$ has unit long-term power.
Therefore, ${{\bf w}}_{il}$ only specifies the precoding direction, whereas the power allocation coefficient $\rho_{il}$ controls the power.
In the CF mMIMO literature, the precoding vectors $\{ {\bf w}_{kl}: l = 1,\ldots,L, k = 1,\ldots,K\}$ are normally selected to match with the uplink combining vectors  as
\begin{equation}\label{eq:ul dl duality of precoding}
    {\bf w}_{kl} = {\bf v}_{kl} = \frac{\bar{\bf v}_{kl}}{\sqrt{{\mathbb E}\{ \| \bar{\bf v}_{kl} \|^2 \}}}.
\end{equation}
This can be motivated by uplink-downlink duality \cite{bjornson2020scalable} and we will derive a similar result below.

The received signal ${y}_{k}^{\rm{dl}} \in \mathbb{C}$ at UE $k$ is
\begin{equation}\label{eq:received signal at ue k}
\begin{aligned}
	&{y}_{k}^{\rm{dl}} = \sum_{l=1}^{L} {\bf h}_{kl}^{\Htran} {\bf x}_{l} + {n}_{k}\\
&= \sum_{l=1}^{L} {\bf h}_{kl}^{\Htran} \sqrt{\rho_{kl}}{\bf w}_{kl} \varsigma_{k} + \sum_{i=1,i\ne k}^{K} \left( \sum_{l=1}^{L} {\bf h}_{kl}^{\Htran} \sqrt{\rho_{il}}{\bf w}_{il} \varsigma_{i}\right) + {n}_{k},
\end{aligned}
\end{equation}
where ${n}_{k} \sim {\cal N}_{\mathbb C} (0,\sigma^2)$ is the independent receiver noise.

\begin{algorithm}[t!]
\label{algo:gw}
\caption{Algorithm for solving ${\sf P}^{\rm gw}$}
\KwIn{${\underline{\bf \Delta}}\in {\mathbb R}^{2KL \times 2KL}$, ${\bf a} \in {\mathbb C}^{KL}$, ${\underline{\pmb \xi}} \in {\mathbb R}^{2KL}$, $\gamma$, $\lambda$, $\mu$, ${n}_{\max}$}

\KwOut{${\bf a} \in {\mathbb C}^{KL}$}

Compute ${\bf X} \in {\mathbb R}^{2KL \times 2KL}$ such that ${\bf X}^{\Ttran} {\bf X} = {\underline{\bf \Delta}}$;\\
${\bar{\pmb \xi}} = ({\bf X}^{\Ttran})^{-1} {\underline{\pmb \xi}}$;\\
\For{$l = 1,\ldots,L$}
{
Extract ${\pmb a}_l$ from ${\bf a}$ as
$
{\pmb a}_l = [a_{1l},\ldots,a_{Kl}]^{\Ttran} \in {\mathbb C}^K
$ such that ${\pmb x}_l = [\Re( {\pmb a}_l )^{\Ttran},  \Im( {\pmb a}_l )^{\Ttran}]^{\Ttran}\in {\mathbb R}^{2K}$
and extract the corresponding ${\bf X}_l \in {\mathbb R}^{2KL \times 2K}$ from ${\bf X}$;\\
Compute the partial residual
$
{\bf r}_l = {\bar{\pmb \xi}} - \sum_{j\ne l} {\bf X}_j {\pmb x}_j \in {\mathbb R}^{2KL}
$;\\
$n = 1$; \\
${{\pmb x}}_l^{-} \leftarrow {{\pmb x}}_l$; \\

\Repeat{$n = n_{\max}$ or convergence}
{
${{\hat{\pmb x}}}_l = {{\pmb x}}_l + \frac{n-1}{n+2} ({{\pmb x}}_l - {{\pmb x}}_l^{-})$;\\
Compute ${\rm Prox}_{\mu,\Omega{'}} ( G( {\hat{{\pmb x}}}_l ) )$ with the help of \eqref{eq:prox_l1+l2}, \eqref{eq:prox l1}, and \eqref{eq:prox l2};\\
${{\pmb x}}_l \leftarrow {\rm Prox}_{\mu,\Omega{'}} ( G( {\hat{{\pmb x}}}_l ) )$;\\
${{\pmb x}}_l^{-} \leftarrow {{\pmb x}}_l$;\\
$n \leftarrow n + 1$;\\
}
Update ${\bf a}$ by replacing the elements indexed by $l$;
}
Obtain ${\bf a} \in {\mathbb C}^{KL}$ with the inverse transformation in \eqref{eq:penalty gw}.
\end{algorithm}

Using the combining vectors from the uplink as in \eqref{eq:ul dl duality of precoding},
$
{\bf g}_{ik} = [{\bf w}_{i1}^{\Htran} {\bf h}_{k1},  \ldots, {\bf w}_{iL}^{\Htran} {\bf h}_{kL}]^{\Ttran} \in {\mathbb C}^L
$ represents
the precoded channels to UE $k$ when the APs transmit to UE $i$.
We define the vector whose elements are the square roots of the power coefficients that the different APs assign to UE $k$  as
\begin{equation}\label{eq:LSFP}
{\bf b}_k = [\sqrt{\rho_{k1}},\ldots,\sqrt{\rho_{kL}}]^{\Ttran} = \sqrt{\rho_{k}} {\pmb \omega}_k \in {\mathbb C}^L
\end{equation}
as the {\it LSFP vector} of UE $k$, where $\rho_k$ is the total transmit power for UE $k$ and ${\pmb \omega}_k = [\omega_{k1},\ldots,\omega_{kL}]^{\Ttran}$ is a unit-norm vector with non-negative entries indicating how the power is allocated among the APs, which is the main concern of this paper.
Notice that $\{{\bf b}_k: k=1,\ldots,K\}$ can be optimized by the CPU in a network-wide manner to maximize certain utilities only employing channel statistics, which is why it is called LSFP.

We can rewrite the received signal at UE $k$ in \eqref{eq:received signal at ue k} as
\begin{equation}\label{eq:received signal at ue k1}
	{y}_{k}^{\rm{dl}} = {\bf g}_{kk}^{\Htran} {\bf b}_k \varsigma_{k} + \sum_{i=1,i\ne k}^{K} {\bf g}_{ik}^{\Htran} {\bf b}_i \varsigma_{i} + {n}_{k},
\end{equation}
where $\{{\bf g}_{ik}^{\Htran} {\bf b}_i:i=1,\ldots,K\}$ represent the effective downlink channels.
We can now compute an achievable downlink SE at UE $k$ by utilizing the hardening bound \cite[Thm. 4.6]{bjornson2017massive}, as
\begin{equation}\label{eq:se dl}
{\sf SE}^{\rm dl}_k = \frac{\tau_{\rm d}}{\tau_{\rm c}} {\log _2}\left( {1 + {\sf SINR}^{\rm dl}_k} \right) \quad \textrm{bit/s/Hz}
\end{equation}
where the effective downlink SINR is given by \cite[Cor. 6.3]{cellfreebook}
\begin{equation} \label{eq:downlink sinr}
{\sf SINR}^{\rm dl}_k = \frac{|{\bf b}_k^{\Htran} \mathbb{E}\{{\bf g}_{kk}\} |^2}{ \sum_{i=1}^K \mathbb{E}\{ | {\bf b}_i^{\Htran} {\bf g}_{ik} | ^2\} - |{\bf b}_k^{\Htran} \mathbb{E}\{{\bf g}_{kk}\} |^2 + \sigma^2}.
\end{equation}
Note that the SE holds for any local precoding and LSFP vectors.
One important difference from LSFD in the uplink is that the SINR in \eqref{eq:downlink sinr} is not a generalized Rayleigh quotient with respect to the LSFP vectors.
As can be seen from \eqref{eq:downlink sinr}, the downlink SINR of a generic UE $k$ is not only affected by the LSFP vector ${\bf b}_k$, but also by all other vectors, i.e., $\{ {\bf b}_i: i=1,\ldots,K \}$. Hence, it is not possible to obtain the optimal LSFP weights to maximize one UE's SE without affecting the others.
The LSFP vectors should be optimized for a certain utility maximization and in general obtaining closed-form results is not possible.

To aid in identifying suitable LSFP vectors, we will establish a novel uplink-downlink duality between the LSFD and LSFP vectors by extending the approach in \cite[Prop.~4]{bjornson2020scalable} that considers  the duality between centralized combining and precoding vectors based on the channel estimates.

\begin{lemm}\label{lemm:duality lsfp}
Consider an uplink system with a set of normalized uplink combining vectors and uplink power coefficients $p_k$, for $k=1,\ldots,K$. Let  $\{ {\widetilde{\bf a}}_k: k=1,\ldots,K \}$ be the unit-norm LSFD weighting vectors. If the LSFP weighting vectors in downlink are selected as
\begin{equation} \label{eq:V-LSFP}
{\widetilde{\bf b}}_k = \sqrt{\rho_k} {\widetilde{\bf a}}_k,
\end{equation}
and the local precoding vectors are selected identically to the normalized uplink combining vectors as in \eqref{eq:ul dl duality of precoding}, then each UE can achieve the same downlink SINR as its uplink SINR ${\widetilde {\sf SINR}}^{\rm ul}_k$. More precisely,
\begin{equation} \label{eq:virtual uplink sinr}
{\sf SINR}^{\rm dl}_k = {\widetilde {\sf SINR}}^{\rm ul}_k = \frac{|{\widetilde{\bf a}}_k^{\Htran} {\pmb \xi}_k |^2} { {\widetilde{\bf a}}_k^{\Htran} ({\bf \Delta}_k - {\pmb \xi}_k {\pmb \xi}_k^{\Htran} ) {\widetilde{\bf a}}_k },\ k = 1,\ldots,K,
\end{equation}
for a certain power allocation policy $\{ \rho_k: k=1,\ldots,K \}$ that satisfies $\sum_{k=1}^K\rho_k\leq \sum_{k=1}^Kp_k$, where ${\bf \Delta}_k$ and ${\pmb \xi}_k$ are given as in \eqref{eq:Delta}-\eqref{eq:xi}.
\end{lemm}
\begin{IEEEproof}
The proof follows the same approach as in \cite{rashid1998transmit}, but for the long-term LSFP and LSFD vectors.
The details are relegated to Appendix \ref{appe:duality} for completeness.
\end{IEEEproof}

Lemma \ref{lemm:duality lsfp}  guarantees that equal effective SINRs can be achieved in the uplink and downlink, if the power allocation coefficients are selected in a unique manner, and the LSFD and LSFP vectors are identical.
This implies that if we optimize the LSFD weights properly, which we have already studied how to do, we can use the same solution for LSFP.
In particular, the sparse LSFD design turns into a sparse LSFP design that provides joint AP-UE assignment and downlink power allocation.
There is only one caveat: the downlink power allocation suggested by the duality result might not comply with the per-AP transmit power constraints.
This can be settled by appropriate centralized downlink power allocation schemes (i.e., selecting the proper per-AP power coefficients $\{\rho_k\}$), elaborated later in this section.
Moreover, since the LSFD and LSFP vectors are computed at the CPU based on the long-term channel statistics and regarded as quasi-static for many time-frequency coherence blocks, the practical fronthaul links would be able to support our proposed distributed processing schemes.

Note that the uplink effective SINR in \eqref{eq:virtual uplink sinr} is a generalized Rayleigh quotient with respect to ${\widetilde{\bf a}}_k$ and, thus, allows computing the LSFD vector ${\widetilde{\bf a}}_k^{\rm opt}$ that maximizes ${\widetilde {\sf SINR}}^{\rm ul}_k$ as in \eqref{eq:O-LSFD}, i.e.,
\begin{equation} \label{eq:V-LSFD}
{\widetilde{\bf a}}_k^{\rm opt} =  {\widetilde c}_k {\bf \Delta}_k^{-1} {\pmb \xi}_k
\end{equation}
where ${\widetilde c}_k \in {\mathbb C}$ being an arbitrary non-zero scaling coefficient.
Then according to \eqref{eq:V-LSFP}, we have
\begin{equation} \label{eq:V-LSFP1}
{\widetilde{\bf b}}_k = \sqrt{\rho_k} \frac{{\widetilde{\bf a}}_k^{\rm opt}}{\|{\widetilde{\bf a}}_k^{\rm opt}\|_2},\ k = 1,\ldots,K,
\end{equation}
which are referred as the {\it V-LSFP} vectors since we use the optimized LSFD vectors but apply a good but heuristic power allocation.

Consequentially, the virtual uplink MSE of UE $k$ becomes
\begin{equation}\label{eq:virtual uplink mse}
    {\widetilde{\sf MSE}}_k^{\rm ul} = {\widetilde{\bf a}}_k^{\Htran} {\bf \Delta}_k {{\widetilde{\bf a}}_k} - 2\sqrt{p_k}\Re \left( {{\widetilde{\bf a}}_k^{\Htran} {\pmb \xi}_k } \right) + p_k
\end{equation}
which is minimized by the virtual LSFD vector in \eqref{eq:V-LSFD} with ${\widetilde c}_k = \sqrt{p_k}$ and, thus, implies that the virtual LSFD vector ${\widetilde{\bf a}}_k^{\rm opt}$ minimizes ${\widetilde{\sf MSE}}_k^{\rm ul}$ as
\begin{equation} \label{eq:minimize virtual uplink mse k}
{\widetilde{\bf a}}_k^{\rm opt} = \argmin{{\widetilde{\bf a}}_k \in {\mathbb C}^{L}} \ {\widetilde{\sf MSE}}_k^{\rm ul}.
\end{equation}

Similar to the uplink MSE in \eqref{eq:uplink mse}, the virtual uplink MSEs also only depend on the UE's own virtual LSFD vector ${\widetilde{\bf a}}_k$, which means one can find the optimal collective virtual LSFD vector
\begin{equation}\label{eq:minimize virtual uplink sum mse}
    {\widetilde{\bf a}}^{\rm opt} = \argmin{{\widetilde{\bf a}} \in {\mathbb C}^{KL}} \ \sum_{k = 1}^K {\widetilde{\sf MSE}}_k^{\rm ul}
\end{equation}
that minimizes the virtual uplink sum MSE of all UEs as
$
    {\widetilde{\bf a}}^{\rm opt} = [ ({\widetilde{\bf a}}_1^{\rm opt})^{\Ttran}, \ldots ,({\widetilde{\bf a}}_K^{\rm opt})^{\Ttran}]^{\Ttran} \in \mathbb{C}^{KL}
$
where $\{ {\widetilde{\bf a}}_k^{\rm opt}: k = 1,\ldots,K \}$ are obtained by simultaneously solving \eqref{eq:minimize virtual uplink mse k}.

\subsection{Centralized Downlink Power Allocation}

Recall from \eqref{eq:LSFP} that $\omega_{kl}$ indicates the fraction of $\rho_k$ that will be sent from AP $l$.
Hence, the power constraint at AP $l$ is be formulated as
\begin{equation}\label{eq:}
\sum_{k\in {\cal D}_l} \rho_k \left| \omega_{kl} \right|^2 \le \rho_{\max},
\end{equation}
where $\rho_{\max}$ is the maximal transmit power of an AP.
Given $\{ \omega_{kl} \}$ are already determined, the algorithms for centralized downlink power allocation can be found in \cite{cellfreebook}.
One good scalable option satisfying the per-AP transmit power constraints, where the computational complexity
does not grow with the number of UEs, is given as \cite{cellfreebook}
\begin{equation}\label{eq:centralized power allocation}
\rho_k = \rho_{\max} \frac{\left( \sum_{l \in {\cal M}_k} \beta_{kl}^{\vartheta} \right)^\kappa \varpi_k^{-\mu}} {\max_{j \in {\cal M}_k} \sum_{i \in {\cal D}_j} \left( \sum_{l \in {\cal M}_i} \beta_{il}^{\vartheta} \right)^\kappa \varpi_i^{1-\mu}},
\end{equation}
where we reshape $\beta_{kl}$ with the exponent $\vartheta$ and
$
\varpi_i = \max_{l\in {\cal M}_i}  | {\omega}_{il} |^2
$
is the largest fraction of $\rho_i$ that any of the serving APs can be assigned to transmit (see \eqref{eq:LSFP}), exponent $\kappa \in \left[{-1,1}\right]$ determines the downlink power allocation behavior, and exponent $\mu \in \left[{0,1}\right]$ is an additional parameter that reshapes the ratio of power allocation between different UEs.
The rationale behind \eqref{eq:centralized power allocation} is that $\rho_k  \propto \left( \sum_{l \in {\cal M}_k} \beta_{kl}^{\vartheta} \right)^\kappa \varpi_k^{-\mu}$, which implies each serving AP of UE $k$ should manage its power constraint as if it transmits with power $\rho_k \varpi_k^{\mu}$.

\renewcommand\arraystretch{1.5}
\begin{table*}[t!]
  \centering
  \fontsize{9}{11}\selectfont
  \caption{System Parameters. }
  \label{tab:paremeter}
    \begin{tabular}{|p{1.45cm}<{\centering}|p{2.2cm}<{\centering}|p{1.6cm}<{\centering}|p{1.2cm}<{\centering}|p{1.45cm}<{\centering}|p{1.75cm}<{\centering}|p{1.65cm}<{\centering}|p{1.75cm}<{\centering}|}
    \hline
        \bf Parameters  & \bf Values  & \bf Parameters  & \bf Values & \bf Parameters  & \bf Values  & \bf Parameters  & \bf Values   \cr\hline
    \hline
        \cellcolor{gray!30}$B$, $\tau_{\rm c}$, $\tau_{\rm p}$  & 20\,MHz, 200, 10  & \cellcolor{gray!30}$\eta_{\rm ue}$, $\eta_{\rm ap}$  & 0.4, 0.4  &
        \cellcolor{gray!30}$\theta$, $\nu$ & 0.5, 0.5  & \cellcolor{gray!30}$\vartheta$, $\kappa$, $\mu$  & 0.2, -0.4, 0.5 \cr\hline
        \cellcolor{gray!30}$p_{\rm p}$, $p_{\max}$  & 0.1\,W, 0.1\,W  & \cellcolor{gray!30}$\rho_{\max}$ & 1\,W  &
        \cellcolor{gray!30}$P_{\rm cpu}^{\rm fix}$, $P_l^{\rm fix}$  & 5\,W, 0.825\,W  & \cellcolor{gray!30}$P_k^{\rm c,ue}$, $P_l^{\rm c,ap}$ & 0.1\,W, 0.2\,W  \cr\hline
        \cellcolor{gray!30}$P_l^{\rm sig}$  & 0.01\,W  & \cellcolor{gray!30}$P_l^{\rm pro}$ & 0.8\,W &
        \cellcolor{gray!30}$P_{\rm cpu}^{\rm dec}$ & 0.8\,W/(Gbit/s) &  \cellcolor{gray!30}$P_{\rm cpu}^{\rm cod}$ & 0.1\,W/(Gbit/s)   \cr\hline
    \end{tabular}
\end{table*}

\subsection{Sparse Optimization for the Downlink}

Similar to the uplink, all values of LSFP vectors obtained by Lemma \ref{lemm:duality lsfp} are non-zero. With the observation that ${\widetilde{\sf MSE}}_k^{\rm ul}$ in \eqref{eq:virtual uplink mse} also possesses the quadratic structure in terms of the virtual LSFD vector ${{\widetilde{\bf a}}_k}$, the sparse algorithms developed in Section \ref{sec:problem} can also applied to enforce sparsity on the the LSFP vectors in the downlink.

\section{Power Consumption Model and Energy Efficiency}\label{sec:power}

The benefit of the proposed sparsity approach to compute the AP-UE association is that we can achieve almost the same SEs as when all APs serve all UEs, but with vastly less fronthaul signaling and signal processing complexity. In this section, we will define a generic power consumption model that can quantify these benefits. The model captures the following main components: a) the radio site power consumption including the power consumed at the UEs $\{P_k^{\rm ue}:\forall k\}$, the active APs $\{P_l^{\rm ap}:\forall l\}$, and fronthaul connections $\{P_l^{\rm fh}:\forall l\}$; and b) the CPU power consumption $P_{\rm cpu}$.
The total power consumption is modeled as
\begin{equation}
 P_{\rm tot} = \sum_{k = 1}^K P_k^{\rm ue} + \sum_{l = 1}^L P_l^{\rm ap} + \sum_{l = 1}^L P_l^{\rm fh} + P_{\rm cpu}.
\end{equation}
We will now model each of these terms in detail.

The power consumption at a generic UE $k$ is
\begin{equation}
 P_k^{\rm ue} = P_k^{\rm c,ue} + \frac{\tau_{\rm p} p_{\rm p} + \tau_{\rm u} p_k}{ \tau_{\rm c} \eta_{\rm ue}}
\end{equation}
where $P_k^{\rm c,ue}$ is the internal circuit power and the second term includes the power consumption of uplink transmission, where $p_{\rm p}$ is the uplink pilot transmit power, $p_k$ is the uplink data transmit power of UE $k$, and $0 < \eta_{\rm ue} \le 1$ is the power amplifier efficiency at the UEs.
${\tau_{\rm p}} / {\tau_{\rm c}}$ and ${\tau_{\rm u}} / {\tau_{\rm c}}$ denotes the fractions of uplink pilot and uplink data transmission, respectively.

The power consumption related to AP $l$ is
\begin{equation}
P_l^{\rm ap} = N P_l^{\rm c,ap} + N |{\cal D}_l|\cdot P_l^{\rm pro} + \frac{\tau_{\rm d}}{ \tau_{\rm c} \eta_{\rm ap}} \sum_{k \in {\cal D}_l} \rho_{kl},
\end{equation}
where $P_l^{\rm c,ap}$ is the internal circuit power per AP antenna, $P_l^{\rm pro}$ is the consumed power for processing the received/transmitted signal of each UE in ${\cal D}_l$, $\rho_{kl}$ is the downlink data transmit power that AP $l$ allocates to UE $k$, and $0 < \eta_{\rm ap} \le 1$ is the power amplifier efficiency at the APs.

The fronthaul connections are used to transfer signals between the APs and the CPU.
The power consumption of each fronthaul link is
\begin{equation}
 P_l^{\rm fh} = P_l^{\rm fix} + \frac{\tau_{\rm u} + \tau_{\rm d}}{ \tau_{\rm c}} |{\cal D}_l|\cdot P_l^{\rm sig},
\end{equation}
where  $P_l^{\rm fix}$ is the fixed power consumption and remaining part describes the load-dependent uplink and downlink signaling, where $P_l^{\rm sig}$ is the signaling power per UE.

The CPU is responsible for processing the signals of all UEs, with power consumption
\begin{equation}
%\begin{aligned}
 P_{\rm cpu} = P_{\rm cpu}^{\rm fix}
 %+ \sum_{k =1}^{K } |{\cal M}_k| \cdot \left( \frac{\tau_{\rm u}}{ \tau_{\rm c}} P_{\rm cpu}^{\rm lsfd} + \frac{\tau_{\rm d}}{ \tau_{\rm c}} P_{\rm cpu}^{\rm lsfp} \right)
  + B \sum_{k =1}^{K } \left( {{\sf{SE}}_k^{\rm ul}} \cdot P_{\rm cpu}^{\rm dec} + {{\sf{SE}}_k^{\rm dl}} \cdot P_{\rm cpu}^{\rm cod} \right)
%\end{aligned}
\end{equation}
where $P_{\rm cpu}^{\rm fix}$ is the fixed power consumption, $B$ is the system bandwidth, $P_{\rm cpu}^{\rm dec}$ is the energy consumption per bit for the final decoding at the CPU, and $P_{\rm cpu}^{\rm cod}$ is the energy consumption per bit for the initial encoding at the CPU.
Typical values for these parameters are given in Table \ref{tab:paremeter}.

With the defined power consumption model, the total EE (in bit/Joule) considering both uplink and downlink is given as \cite{bjornson2017massive,ngo2017total}
\begin{equation}\label{eq:ee}
 {\sf{EE}} = {B \cdot \sum_{k = 1}^K  \left( {\sf{SE}}_k^{\rm ul} + {\sf{SE}}_k^{\rm dl} \right) }/{P_{\rm tot}}.
\end{equation}
    
\renewcommand\arraystretch{1.5}
\begin{table*}[t!]
  \centering
  \fontsize{9}{11}\selectfont
  \caption{The Schemes and Benchmarks for the Uplink. }
  \label{tab:scheme uplink}
    \begin{tabular}{|p{1.8cm}<{\centering}|p{3.2cm}<{\centering}|p{10cm}|}
    \hline
      \bf Schemes  & \bf AP-UE association  & {\bf LSFD}: ${\bf a}_k = [a_{k1},\ldots,a_{kL}]^{\Ttran},\ k = 1,\ldots,K$ \cr\hline
    \hline
      \cellcolor{gray!30}O-LSFD \cite{nayebi2016performance} & All APs serve all UEs. & ${\bf a}_k$ is optimized by \eqref{eq:O-LSFD}. \cr\hline
      \cellcolor{gray!30}P-LSFD \cite{cellfreebook} & Heuristic scheme \cite{bjornson2020scalable}
      & ${\bf a}_k = c_k \left( \sum_{i\in {\cal P}_k} p_i {\mathbb E} \{{\bf g}_{ki} {\bf g}_{ki}^{\Htran}\} + \sigma^2 \vect{I}_L \right)^{-1} {\pmb \xi}_k$

      where ${\cal P}_k = \left\{ i: {\cal M}_i \cap {\cal M}_k \ne \emptyset,\ i = 1,\ldots,K\right\}$. \cr\hline
      \cellcolor{gray!30}S-LSFD & \multicolumn{2}{l|}{\makecell[l]{\specialrule{0em}{3pt}{0em}Sparse optimization in \eqref{eq:p0} is utilized to enforce sparsity on ${\bf a}_k$ achieved from scheme {O-LSFD}.}} \cr\hline
    \end{tabular}
\end{table*}

\renewcommand\arraystretch{1.5}
\begin{table*}[t!]
  \centering
  \fontsize{9}{11}\selectfont
  \caption{The Schemes and Benchmarks for the Downlink. }
  \label{tab:scheme downlink}
    \begin{tabular}{|p{1.3cm}<{\centering}|p{3cm}<{\centering}|p{10.7cm}|}
    \hline
        \bf Schemes  & \bf AP-UE association  & {\bf LSFP}: ${\bf b}_k = [\sqrt{\rho_{k1}},\ldots,\sqrt{\rho_{kL}}]^{\Ttran}$, $\ k = 1,\ldots,K$ \cr\hline
    \hline
    \cellcolor{gray!30}FPA \cite{interdonato2019scalability} & Heuristic scheme \cite{bjornson2020scalable}
      & ${\bf b}_k = [\sqrt{\rho_{k1}},\ldots,\sqrt{\rho_{kL}}]^{\Ttran}$, where $\{\rho_{kl} \}$ are selected according to \eqref{eq:distributed power allocation}. \cr\hline
      \cellcolor{gray!30}H-FPA & Heuristic scheme \cite{bjornson2020scalable}
      & ${\bf b}_k = \sqrt{\rho_k} \frac {{\pmb \rho}_k}  {\| {\pmb \rho}_k \|_2}$, where ${\pmb \rho}_k = [\rho_{k1}, \ldots, \rho_{kL}]^{\Ttran}$, $\{\rho_{kl} \}$ are selected according to \eqref{eq:distributed power allocation}, and $\{\rho_{k} \}$ are selected according to  \eqref{eq:centralized power allocation}.\cr\hline
      \cellcolor{gray!30}V-LSFP & All APs serve all UEs.
      & ${\bf b}_k$ is computed by \eqref{eq:V-LSFP1} where $\{\rho_{k}\}$ are selected according to  \eqref{eq:centralized power allocation}. \cr\hline
      \cellcolor{gray!30}P-LSFP & Heuristic scheme \cite{bjornson2020scalable}
      & ${\bf b}_k = \sqrt{\rho_k} \frac{{\widetilde{\bf a}}_k}{\| {\widetilde{\bf a}}_k \|_2}$,
      where ${\widetilde{\bf a}}_k =  {\widetilde c}_k \left( \sum_{i\in {\cal P}_k} {\mathbb E} \{{\bf g}_{ki} {\bf g}_{ki}^{\Htran}\} + \sigma^2 {\bf I}_L \right)^{-1} {\pmb \zeta}_k$
      and $\{\rho_{k}\}$ are selected according to \eqref{eq:centralized power allocation}.\cr\hline
      \cellcolor{gray!30}S-LSFP & \multicolumn{2}{l|}{\makecell[l]{\specialrule{0em}{3pt}{0em}Sparse optimization in \eqref{eq:p0} is utilized to enforce sparsity on ${\bf b}_k$ achieved from scheme {V-LSFP}.}} \cr\hline
      \cellcolor{gray!30}SV-LSFP & Association achieved from scheme {S-LSFD}
      & ${\bf b}_k$ is computed by \eqref{eq:V-LSFP1} where $\{\rho_{k}\}$ are selected according to  \eqref{eq:centralized power allocation}. \cr\hline
    \end{tabular}
\end{table*}

\section{Numerical Results}\label{sec:results}

In this section, we quantify the performance achieved by our proposed LSF processing schemes in Section \ref{sec:uplink} and Section \ref{sec:downlink}, considering different combining and precoding schemes and AP deployment setups. Specifically, the L-MMSE and MR combiners are used for the uplink, and the L-MMSE and MR precoders are used for the downlink. We will measure performance in terms of SE, EE, and number of serving APs per UE (marked as ``no. AP/UE" in the figures).

We consider two different AP deployments: a) $L = 40$ APs with $N = 4$ antennas and b) $L=160$ APs with $N=1$ antenna.
The total number of antennas is $LN=160$ in both cases.
All APs and $K = 20$ UEs are distributed in the coverage area of $0.5\times 0.5$ km$^2$ at random following an independent and uniform distribution.
We use the wrap-around topology to approximate an infinitely large network. % where all UEs are effectively in the center of the simulation area.% and are subject to the interference from all directions.
The 3GPP Urban Microcell model \cite{access2017further} is used to compute the large-scale propagation conditions, such as pathloss and shadow fading.
The spatial correlation matrices are generated by using the Gaussian local scattering model with the azimuth and elevation angular standard deviation of $10^\circ$ and $10^\circ$, respectively, as described in \cite[Sec. 2.5.3]{cellfreebook}.
The SE results with L-MMSE combining/precoding are obtained from Monte Carlo simulations, while the results with MR combining/precoding are analytically computed according to the closed-form expressions in \cite[Cor. 2]{bjornson2020scalable}.
After obtaining the SE and AP-UE association results, the EE values were computed using \eqref{eq:ee} with our proposed power consumption model.
Moreover, the convergence of our proposed proximal algorithms in Section
\ref{sec:problem} are validated by comparing them to CVX SDPT3 (Ver. 2.2) \cite{cvx2015}. We use
$\tau_{\rm d} \!= \!0$ and $\tau_{\rm u} \!= \!0$ when evaluating the performance for the uplink and the downlink, respectively.
Unless otherwise specified, all other system parameters are given in Table \ref{tab:paremeter} and originate from \cite{bjornson2019making,ngo2017total,bjornson2015optimal} (and reference therein).

\subsection{Considered Schemes and Benchmarks}\label{subsec:schemes}

In the uplink, the transmit powers $\{p_k : \forall k\}$ are selected according to the fractional power control policy \cite{cellfreebook,chen2020structured}
\begin{equation}\label{eq:power control}
{p_k} = p_{\rm max} \frac{ \min_{i\in \{1,\ldots,K\}}\left(\sum_{l\in {\cal M}_i} \beta_{il}\right)^\theta }{\left(\sum_{l\in {\cal M}_k} \beta_{kl}\right)^\theta},
\end{equation}
where $p_{\rm max}$ is the maximal transmit power of a UE and $\theta \!\in\! \left[{0,1}\right]$ determines the control behavior.
$\theta = 0$ leads to equal power control and $\theta \to 1$ promotes more user fairness.

To demonstrate the performance improvements of our joint AP-UE association and LSFD, we compare the proposed {S-LSFD} with two benchmarks: {O-LSFD} and {partial LSFD (P-LSFD)}.
The details of these benchmarks are summarized in Table \ref{tab:scheme uplink}.

For the downlink, the precoding vectors are computed using \eqref{eq:ul dl duality of precoding}.
The transmit powers can be selected in a distributed manner as \eqref{eq:centralized power allocation} \cite{interdonato2019scalability,bjornson2020scalable}
\begin{equation}\label{eq:distributed power allocation}
{\rho _{kl}} =
{{\rho _{\max}} \frac{{(\beta _{kl})}^\nu}{{\sum_{i \in {{\cal D}_{l}}} {{(\beta _{il})}^\nu}  }}}
\end{equation}
if $k\in {{\cal D}_{l}}$ and otherwise ${\rho _{kl}} = 0$, with $\nu \in \left[{0,1}\right]$ determining the power allocation behavior.
$\nu = 0$ leads to equal power allocation and $\nu \to 1$ allocates more power to the UEs with better channel conditions.
If the directions of the LSFP vectors $\{\vect{b}_k\}$ are already determined, the per-AP power coefficients can be selected in centralized manner as \eqref{eq:centralized power allocation} \cite{cellfreebook}.

To highlight the performance improvements of our V-LSFP using uplink-downlink duality in Lemma \ref{lemm:duality lsfp} and joint AP-UE association and LSFP, we propose several schemes, namely {heuristic FPA (H-FPA)}, {V-LSFP}, {partial LSFP (P-LSFP)}, {S-LSFP}, and {sparse V-LSFP (SV-LSFP)}.
We consider a benchmark where $\{\rho_{kl},\forall k,l\}$ for $\{{\bf b}_k,\forall k\}$ are selected according to \eqref{eq:distributed power allocation}, which is referred to as scheme {FPA} in the numerical results.
These schemes are elaborated in Table \ref{tab:scheme downlink}.

\subsection{Analysis for the Uplink}

\begin{figure}[t!]
\centering
\includegraphics[scale=0.6]{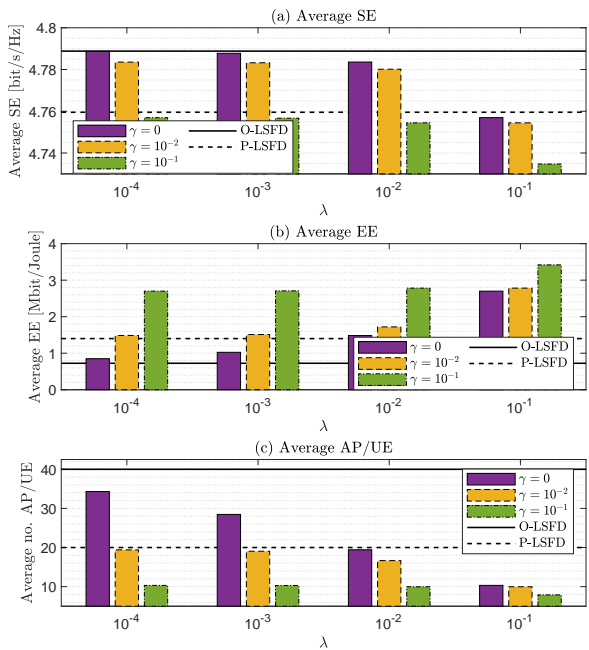}
\caption{Uplink average SE, EE, and number of serving APs per UE with L-MMSE combining ($L=40,N=4$).
\label{fig:ul_mmse_l40}}
\end{figure}

In Fig.~\ref{fig:ul_mmse_l40}, we evaluate the considered performance metrics achieved by L-MMSE combining with the multi-antenna AP setup (i.e., $L=40, N=4$), where the average SE, EE, and number of serving APs per UE are demonstrated in Fig.~\ref{fig:ul_mmse_l40}(a), Fig.~\ref{fig:ul_mmse_l40}(b), and Fig.~\ref{fig:ul_mmse_l40}(c), respectively.
We compare the proposed {S-LSFD} with the benchmarks {O-LSFD} and {P-LSFD} for various values of the regularization parameters $\lambda$ and $\gamma$, where $\gamma \!=\! 0$ stands for the case of EW-sparsity.
The vertical scale intervals are set to emphasize how small/large the gaps are between the curves.
The first observation is that the average SE decreases as $\lambda$ and $\gamma$ increase since the average number of serving APs per UE decreases.
It is clear that although our proposed {S-LSFD} {\it slightly} reduces the SE by around $1\%$ (for large values of $\lambda$ and $\gamma$ that each UE is served by its most essential APs), it significantly increases the EE. There is a $4\times$ EE gain compared to {O-LSFD} where all APs serve all UEs.
Compared to {P-LSFD}, {S-LSFD} provides larger SE and similar EE by using approximately the same number of the serving APs per UE (with $\lambda = 10^{-4}, \gamma = 10^{-2}$) and also provides $1.92\times$ EE and similar SE by using half number of the serving APs per UE (with $\lambda = 10^{-1},\gamma =0$).
The reason for this is that our joint AP-UE association and LSFD design outperform {P-LSFD} where the association and LSFD are performed separately.
That also implies that {S-LSFD} is capable of making a better tradeoff between the SE and EE than {P-LSFD} by adjusting $\lambda$ and $\gamma$.

\begin{figure}[t!]
\centering
\includegraphics[scale=0.6]{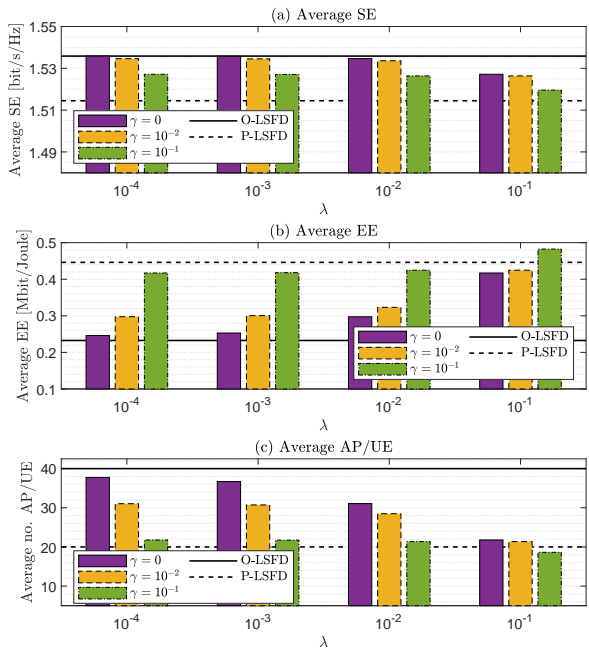}
\caption{Uplink average SE, EE, and number of serving APs per UE with MR combining ($L=40,N=4$).
\label{fig:ul_mr_l40}}
\end{figure}

Fig.~\ref{fig:ul_mr_l40} shows the results achieved by MR combining with the multi-antenna AP setup.
Compared to Fig.~\ref{fig:ul_mmse_l40}, it is clear that L-MMSE combining outperforms MR regarding SE thanks to its interference suppression.
Moreover, although MR may require less processing power than L-MMSE, it still cannot compensate for its disadvantage of throughput, which leads to lower EE.
Similar trends in SE and EE concerning $\lambda$ and $\gamma$ as in Fig.~\ref{fig:ul_mmse_l40} can be observed.
It is worth noting in Fig.~\ref{fig:ul_mmse_l40}(c) and Fig.~\ref{fig:ul_mr_l40}(c) that with the same AP deployment, MR combining benefits more from using many APs, which is reflected by having more serving APs per UE than with L-MMSE combining for all combinations of $\lambda$ and $\gamma$.
This is because there is so much interference when using MR that also APs that have rather weak channels to the UE can positively improve the SE (see the ranges in Fig.~\ref{fig:ul_mmse_l40}(a) and Fig.~\ref{fig:ul_mr_l40}(a)).

\begin{figure}[t!]
\centering
\includegraphics[scale=0.6]{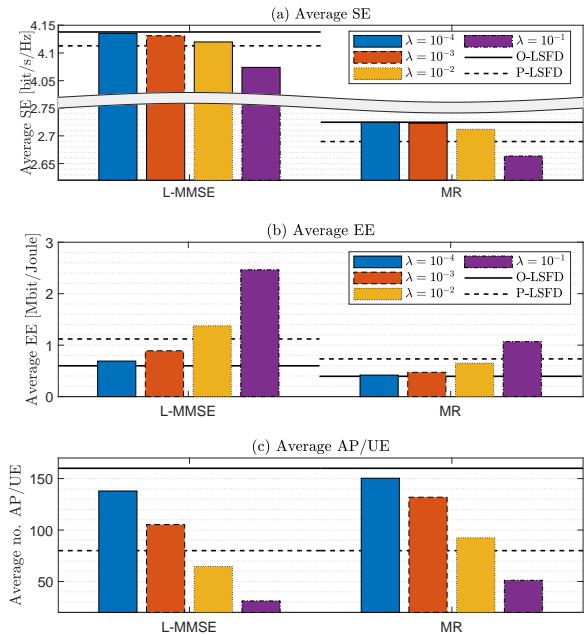}
\caption{Uplink average SE, EE, and number of serving APs per UE with different combiners ($\gamma = 0, L=160,N=1$).
\label{fig:ul_l160}}
\end{figure}

Since the influence of the regularization parameter $\gamma$ is similar to that of $\lambda$, which has been demonstrated in Fig.~\ref{fig:ul_mmse_l40} and Fig.~\ref{fig:ul_mr_l40}, the following figures with respect to sparse optimization will only consider the EW-sparsity (i.e., $\gamma = 0$).
Fig.~\ref{fig:ul_l160} is dedicated to the single-antenna AP setup (i.e., $L=160, N=1$), where L-MMSE and MR combining are both used.
Since the SE gaps between L-MMSE and MR is very large, we break the vertical axis in Fig.~\ref{fig:ul_l160}(a) and remove the unnecessary blank space for clear presentation.
Compared to Fig.~\ref{fig:ul_mmse_l40} and Fig.~\ref{fig:ul_mr_l40}, we notice that the multi-antenna AP setup outperforms the single-antenna AP setup with L-MMSE combining case while it is the opposite with MR combining.
The reason is that in the L-MMSE case, the interference suppression gain enabled by multiple antennas is more beneficial than the macro-diversity gain brought by having more APs.
Conversely, the macro-diversity gain dominates in the MR case, which relies on it for avoiding interference.
Another observation is that the EE gaps between L-MMSE and MR is larger with multi-antenna APs (between Fig.~\ref{fig:ul_mmse_l40}(b) and Fig.~\ref{fig:ul_mr_l40}(b)) than with single-antenna APs (see Fig.~\ref{fig:ul_l160}(b)) thanks to the interference suppression.

\begin{figure}[t!]
\centering
\includegraphics[scale=0.6]{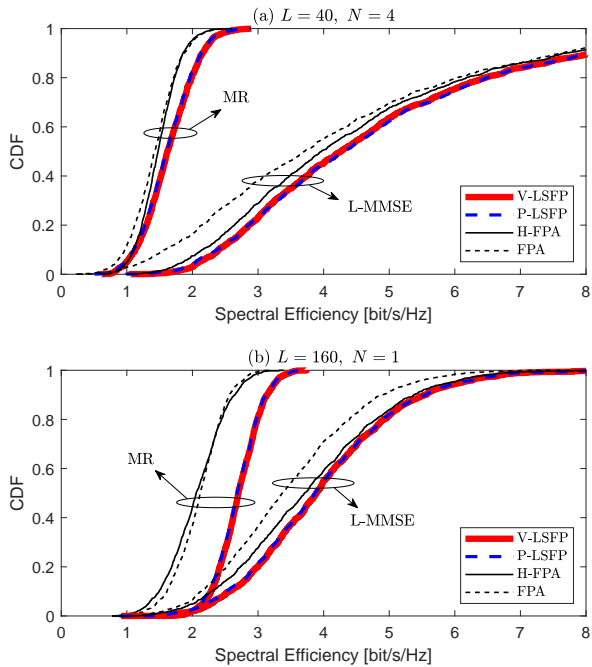}
\caption{Downlink SE per UE of the non-sparse schemes with different precoders and AP deployment setups.
\label{fig:dl_se_cdf}}
\end{figure}

\subsection{Analysis for the Downlink}

According to whether it involves the sparse optimization or not, the schemes considered in the downlink can be divided into two categories: the non-sparse schemes and the sparse schemes.
The former includes {FPA}, {H-FPA}, {V-LSFP}, and {P-LSFP}, and the latter includes {S-LSFP} and {SV-LSFP}.
We first evaluate the SE and EE performance of the non-sparse schemes to highlight the performance improvements achieved by our proposed V-LSFP design.

Fig.~\ref{fig:dl_se_cdf} shows the cumulative distribution function (CDF) of the downlink SE per UE.
The proposed schemes {V-LSFP}, {P-LSFP}, and {H-FPA} are compared to the benchmark {FPA} with L-MMSE and MR precoding and two considered AP deployment setups.
The first observation is {H-FPA} outperforms {FPA} by $1.5\times$ on $95\%$-likely SE thanks to the additional centralized FPA in \eqref{eq:centralized power allocation}.
The $95\%$-likely SE is further improved by {V-LSFP} and {P-LSFP} to around $1.7\times$, which both exploit the uplink-downlink duality proposed in Lemma \ref{lemm:duality lsfp} to design the direction of the LSFP weighting vectors.
The reason is the unit-norm virtual LSFD vectors $\{ {\widetilde {\bf a}}_k \}$ used in {V-LSFP} and {P-LSFP} are optimized in \eqref{eq:V-LSFD} for interference suppression, and, thus, specify the fractions of $\rho_k$ for the serving APs better than {H-FPA}, where the fractions of $\rho_k$ are determined by the distributed PFA in \eqref{eq:distributed power allocation}.
Scheme {P-LSFP} has a slightly lower $95\%$-likely SE compared to {V-LSFP} due to the reduced number of serving APs per UE.
By comparing Fig.~\ref{fig:dl_se_cdf}(a) and Fig.~\ref{fig:dl_se_cdf}(b), we notice that the SE gap between the proposed schemes and the benchmark {FPA} is large with $L=40,N=4$ and shrinks with $L=160,N=1$ in the L-MMSE case, while it is the opposite in the MR case.
This is because the L-MMSE precoder benefits from the interference suppression gain enabled by multiple antennas more than the macro-diversity gain brought by having more APs, and the MR precoder is the opposite.

\begin{figure}[t!]
\centering
\includegraphics[scale=0.6]{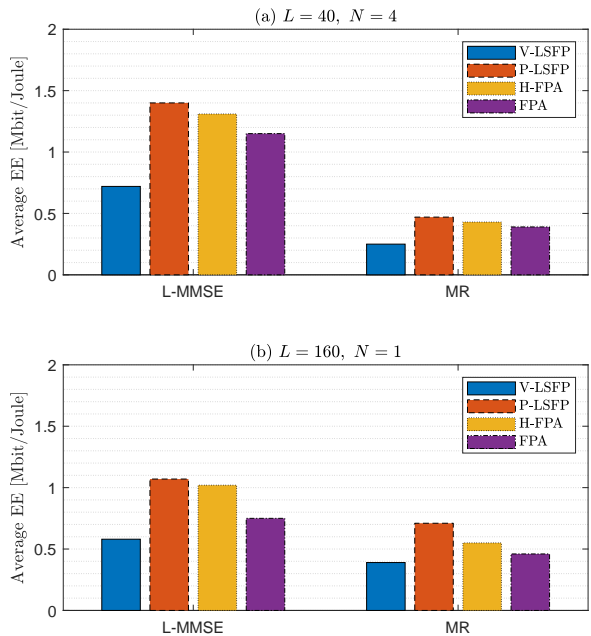}
\caption{Downlink average EE of the non-sparse schemes with different precoders and AP deployment setups.
\label{fig:dl_ee_mean}}
\end{figure}

The average EE of the non-sparse schemes is shown in Fig.~\ref{fig:dl_ee_mean} (with two precoders and two AP deployment setups), from which we observe that {FPA} outperforms {V-LSFP} where all APs serve all UEs.
{P-LSFP} and {H-FPA} achieve higher EE than {FPA} by allocating the downlink transmit power more appropriately.
When comparing the EE gaps between the two AP deployment setups, we have a similar observation of the SE gaps in Fig.~\ref{fig:dl_se_cdf} for the similar reason.

\begin{figure}[t!]
\centering
\includegraphics[scale=0.6]{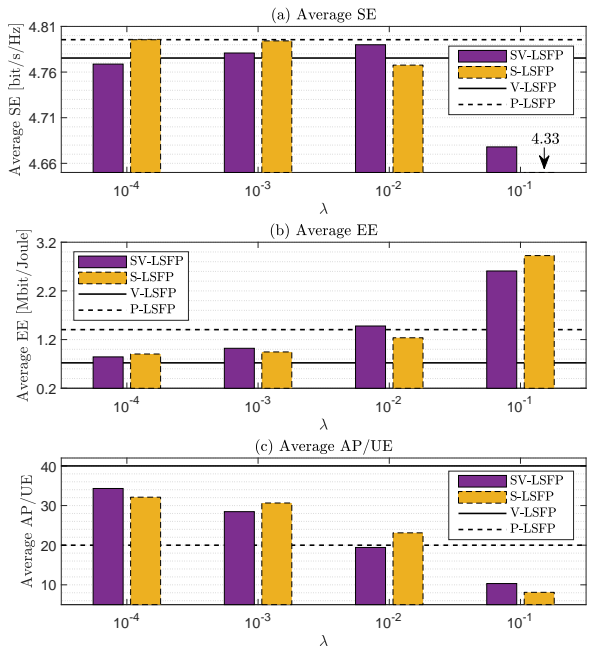}
\caption{Downlink average SE, EE, and number of serving APs per UE with L-MMSE precoding ($\gamma = 0, L=40,N=4$).
\label{fig:dl_mmse_g0_l40}}
\end{figure}

From Fig.~\ref{fig:dl_se_cdf} and Fig.~\ref{fig:dl_ee_mean} it is clear that {V-LSFP} and {P-LSFP} outperform the other two non-sparse schemes on SE and act as the lower and upper bound of the average EE, respectively.
Therefore, to highlight the performance of the sparse LSFP schemes, we only include {V-LSFP} and {P-LSFP} into the following comparisons for concise presentation.
Moreover, as we already observed, the L-MMSE precoder outperforms the MR precoder and benefits more from the multi-antenna AP setup. Thus, only the case with L-MMSE precoding and $L\!=\!40, N\!=\!4$ is presented.

In Fig.~\ref{fig:dl_mmse_g0_l40}, we evaluate the average SE and EE of our LSFP schemes by considering L-MMSE precoding with $L=40,N=4$.
Unlike the uplink case in Fig.~\ref{fig:ul_mmse_l40}, {V-LSFP} has a lower average SE compared to its partial version {P-LSFP}.
One reason for this result is that an appropriate transmit power allocation influenced by the AP-UE association is essential for downlink operation, where the signals from a remote AP might not contribute to the desired signal of the intended UEs, and even cause interference for the other UEs if the transmit power is not well allocated.
 Another reason comes from the suboptimality of the L-MMSE precoding unlike its uplink counterpart.
For the sparse schemes {S-LSFP} and {SV-LSFP}, we observe that although the sparse optimization of {S-LSFP} is directly performed on the downlink {V-LSFP} vectors, it could not maintain an absolute advantage over {SV-LSFP}, which exploits the sparse association from uplink for the downlink operation, on both SE and EE.
In fact, these two sparse LSFP schemes are comparable with each other.
The one with more serving APs per UE might win on SE but lose on EE.
This is because {V-LSFP} is a heuristic scheme where the V-LSFP weighting vectors and the precoding vectors are computed by using uplink-downlink duality.
As a consequence, the improvement of directly performing sparse optimization in the downlink is not guaranteed.
When compared to the non-sparse schemes, {S-LSFP} and {SV-LSFP} are competitive with comparable SE with {P-LSFP}, provide higher EE than {V-LSFP}, and a better tradeoff between the SE and EE.
In addition, we notice that the average SE in Fig.~\ref{fig:dl_mmse_g0_l40}(a) is unimodal with respect to $\lambda$, which implies that there exists a value of $\lambda$ that provides maximum average SE.

\begin{figure}[t!]
\centering
\includegraphics[scale=0.6]{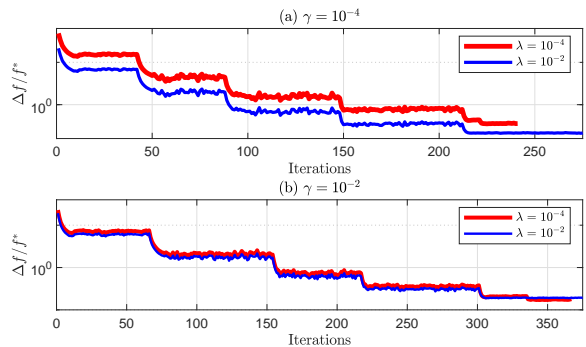}
\caption{Convergence accuracy with different sparsity parameters ($L=40,N=4$).
\label{fig:conv}}
\end{figure}

\begin{figure}[t!]
\centering
\includegraphics[scale=0.6]{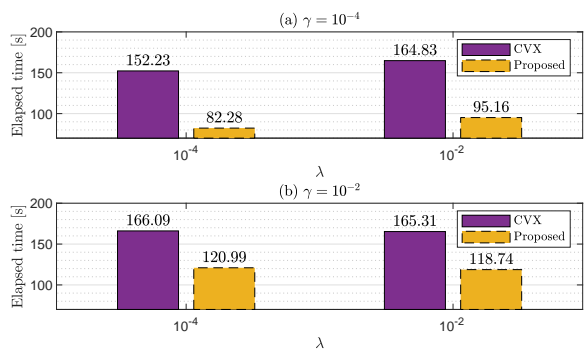}
\caption{Elapsed time for convergence in Fig.~\ref{fig:conv} with CVX and our proposed algorithm in Algorithm \ref{algo:gw} ($L=40,N=4$).
\label{fig:time}}
\end{figure}

\subsection{Algorithmic Convergence}

We consider two metrics to validate the convergence of our proposed proximal algorithm with randomly generated matrices and vectors in the optimization problems: the accuracy  and the elapsed time.
The accuracy  is shown in Fig.~\ref{fig:conv} and is defined as $\Delta f / f^\ast$, which is the function value difference $\Delta f \!=\! f \!-\! f^\ast$ normalized by the ``optimal" value $f^\ast$ obtained by CVX.
The elapsed times for convergence with different sparsity parameters are given in Fig.~\ref{fig:time}, where the CVX solver is considered as the benchmark for our Algorithm \ref{algo:gw}.
Fig.~\ref{fig:conv} validates the correctness of our algorithm by showing the accuracy of $10^{-4}$.
Also, we observe that the proximal algorithm converges faster with a larger $\lambda$, where the staircase comes from the warm-restart operation.
The results in Fig.~\ref{fig:time} demonstrate the effectiveness of our algorithm where the elapsed time of our algorithm is much less than that of CVX, especially when $\lambda$ and $\gamma$ are small.
And this advantage in terms of effectiveness will grow in large-scale networks.

\section{Conclusion}\label{sec:conclusion}%\vspace{-0.5em}

This paper developed a joint optimization framework for the AP-UE association and distributed decoding/precoding in CF mMIMO systems. It is based on formulating and solving two sparsity-inducing MSE-minimizing problems that aim for EW and GW sparsity, respectively.
The former limits the number of UEs served on average by each AP and the latter also encourages APs to not serve any UEs when not essential, both in an effort to reduce signaling and computations to improve the EE.
We developed proximal algorithms to solve the formulated sparsity problems given the predetermined sparsity parameters, where the BCD approach was used for the GW case.
Based on the sparse optimization, we proposed the S-LSFD scheme for the uplink.
For the downlink, we first proposed the new V-LSFP by using uplink-downlink duality, which achieves a good heuristics distributed precoding.
By only considering the UEs with common serving APs during the interference suppression of V-LSFP, we proposed the P-LSFP where each UE is served by a limited number of APs instead all of them.
Then, we proposed the S-LSFP where the sparse association is directly obtained in the downlink, and the SV-LSFP where the association is obtained by S-LSFD in the uplink and then used as a priori for P-LSFP in the downlink.

The numerical results demonstrated that our joint optimization of AP-UE association and signal processing outperforms the existing approach, in which these operations are performed separately. The gain is especially large when using L-MMSE combining with multi-antenna APs.
For example, in the uplink, the proposed S-LSFD achieved $4\times$ higher EE than O-LSFD, while only losing $1\%$ in SE.
S-LSFD achieved a $1.92\times$ EE gain and similar SE by using half number of serving APs per UE.
For the downlink, our H-FPA achieved $1.5\times$ $95\%$-likely SE compared to FPA by using further power allocation (with L-MMSE precoder and multi-antenna APs).
Under the same setup, our V-LSFP and P-LSFP increased this $95\%$-likely SE advantage to $1.7\times$ thanks to the virtual uplink optimization.
When considering EE, FPA outperforms V-LSFP while falling behind P-LSFP and H-FPA, where the former shows higher EE.
The sparse optimization also works well in the downlink where S-LSFP and SV-LSFP achieved comparable SE with P-LSFP, higher EE than V-LSFP, and a better tradeoff between the SE and EE.
The comparison between S-LSFP and SV-LSFP implies that the sparse associations in the uplink and downlink are analogical when the proposed uplink-downlink duality is used.

\begin{appendices}

\section{Proof of Lemma \ref{lemm:prox}}\label{appe:prox}
Since \eqref{eq:p2.4} is convex, the optimal solution ${{\pmb x}}_l^\ast$ is characterized by the subgradient equation
\begin{equation}
G( {{\pmb x}}_l^{n})-{{\pmb x}}_l^\ast = \mu \gamma \partial \| {{\pmb x}}_l^\ast \|_2 + \mu \lambda \partial \|{{\pmb x}}_l^\ast \|_1,
\end{equation}
where
\begin{equation}
\partial \| {{\pmb x}}_l^\ast \|_2 = \begin{cases}
{\frac{{{\pmb x}}_l^\ast}{\| {{\pmb x}}_l^\ast \|_2},} &{{\rm if}\ {{\pmb x}}_l^\ast \ne {\bf 0}}\\
{\in \{ {\bf u} : \| {\bf u}\|_2 \le 1 \},} &{{\rm otherwise}}
\end{cases}
\end{equation}
and
\begin{equation}
[\partial \| {{\pmb x}}_l^\ast \|_1]_i = \begin{cases}
{{\rm sign}([ {{\pmb x}}_l^\ast]_i),} &{{\rm if}\ [ {{\pmb x}}_l^\ast]_i \ne { 0}}\\
{\in \{ {u} : | {u}| \le 1 \},} &{{\rm otherwise}}
\end{cases}, \quad i = 1,\ldots,2K,
\end{equation}
are the subgradients of $\| {{\pmb x}}_l^\ast \|_2$ and $\| {{\pmb x}}_l^\ast \|_1$, respectively.
After some algebraic manipulations, we notice that the subgradient equations are satisfied with ${{\pmb x}}_l^\ast = {\bf 0}$ if
$
\| {\rm Prox}_{\mu \lambda,\ell_1} ( G( {{\pmb x}}_l^{n}) ) \|_2 \le \mu \gamma,
$
and otherwise ${{\pmb x}}_l^\ast$ satisfy
$
\| {\rm Prox}_{\mu \lambda,\ell_1} ( G( {{\pmb x}}_l^{n}) ) \|_2 = {\| {{\pmb x}}_l^\ast \|_2} + \mu \gamma.
$
Then with the definition of the proximal operator of the $\ell_2$-norm in \eqref{eq:prox l2}, we obtain the closed-form expression of ${{\pmb x}}_l^\ast$ as in \eqref{eq:prox_l1+l2} and this concludes the proof of Lemma \ref{lemm:prox}.

\section{Proof of Lemma \ref{lemm:duality lsfp}}\label{appe:duality}

We prove our claim by first noting that the uplink CF SINR given in \eqref{eq:uplink sinr} has the same form as in the uplink SINR in \cite[Eq.~(10)]{rashid1998transmit} when the LSFD vectors take the role of receive beamforming weight vectors. Similarly the downlink CF SINR given in \eqref{eq:downlink sinr} has the same form as in the downlink SINR in \cite[below~Eq.~(14)]{rashid1998transmit} when the normalized LSFP vectors ${\bf b}_k/\sqrt{\rho_k}$ take the role of transmit beamforming weight vectors. Now, consider the virtual uplink system with the SINRs ${\widetilde {\sf SINR}}^{\rm ul}_k$ in \eqref{eq:virtual uplink sinr}. Then, the problem of minimizing total downlink power $\sum_{k=1}^K\rho_k$ under the downlink SINR constraints ${\sf SINR}^{\rm dl}_k\geq{\widetilde {\sf SINR}}^{\rm ul}_k$ is feasible and at the optimal solution, ${\sf SINR}^{\rm dl}_k={\widetilde {\sf SINR}}^{\rm ul}_k$ is achievable in the downlink when the LSFP vectors are selected as in \eqref{eq:V-LSFP}. The optimal objective value is $\sum_{k=1}^K\rho_k\leq \sum_{k=1}^Kp_k$. The equality is achieved when the power coefficients $p_k$ and the LSFD vectors $\widetilde{\bf a}_k$ are the optimal solutions to the uplink power minimization problem, as proved in detail in \cite[p.~1442]{rashid1998transmit}.
\end{appendices}

\bibliographystyle{IEEEtran}
\bibliography{IEEEabrv,ref}

%----------------------------biography----------------------------

\begin{IEEEbiography}[{\includegraphics[width=1in,height=1.25in,clip,keepaspectratio]{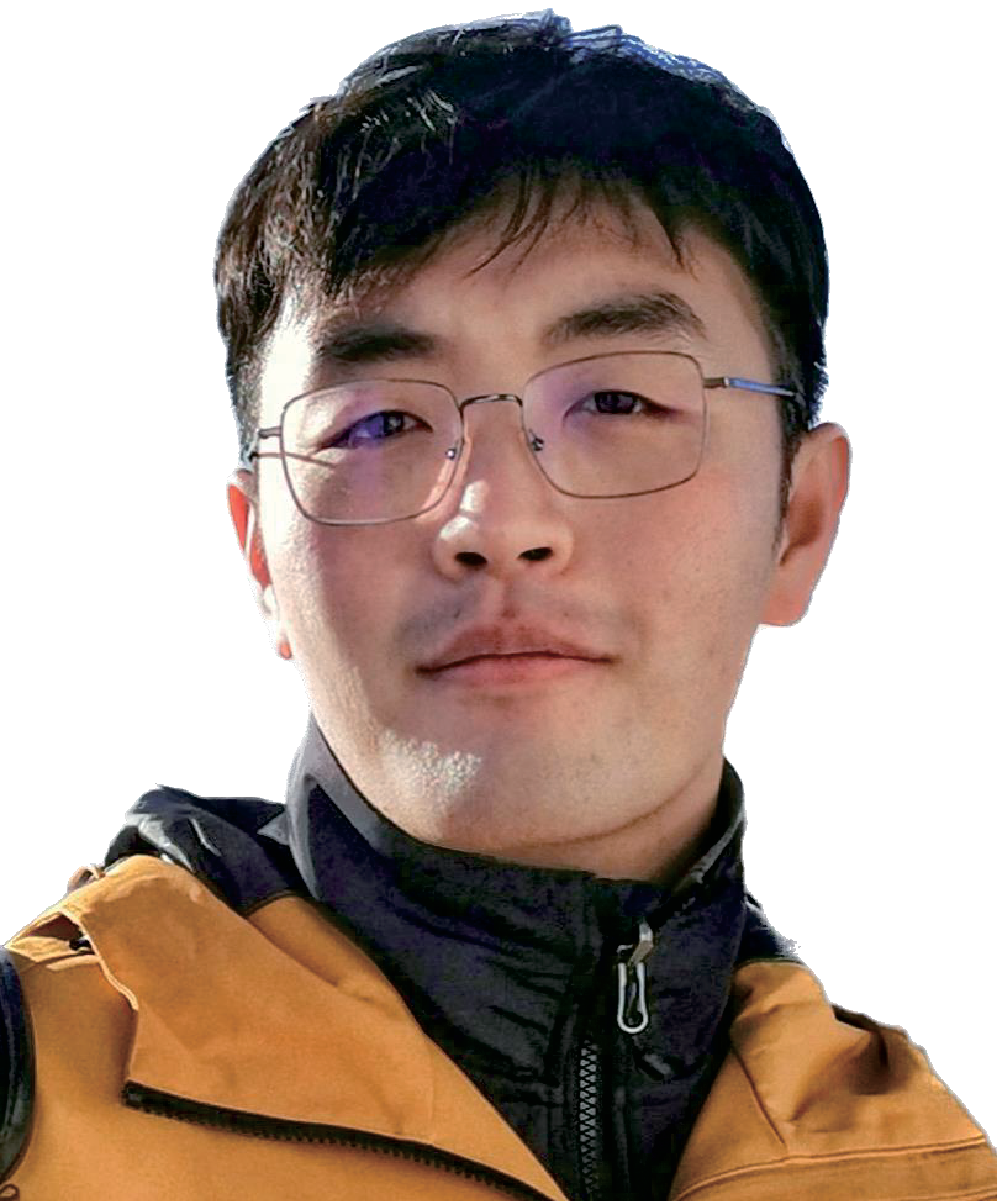}}]
    {Shuaifei Chen} received the B.S. degree in communication engineering from Beijing Jiaotong University, China, in 2018. Since 2018, he is currently a Ph.D. student at Beijing Jiaotong University. From 2019 to 2020, he visited the Department of Communication Systems, Link{\"o}ping University, Sweden. From 2021 to 2022, he visited the Division of Communication Systems, KTH Royal Institute of Technology, Sweden. His research interests include signal processing and resource allocation for wireless communications, cell-free massive MIMO, and electromagnetic information theory for 6G multiple antenna technologies. He was recognized as an Exemplary Reviewer of \textsc{IEEE Transactions on Communications} in 2021.
\end{IEEEbiography}

\begin{IEEEbiography}[{\includegraphics[width=1in,height=1.25in,clip,keepaspectratio]{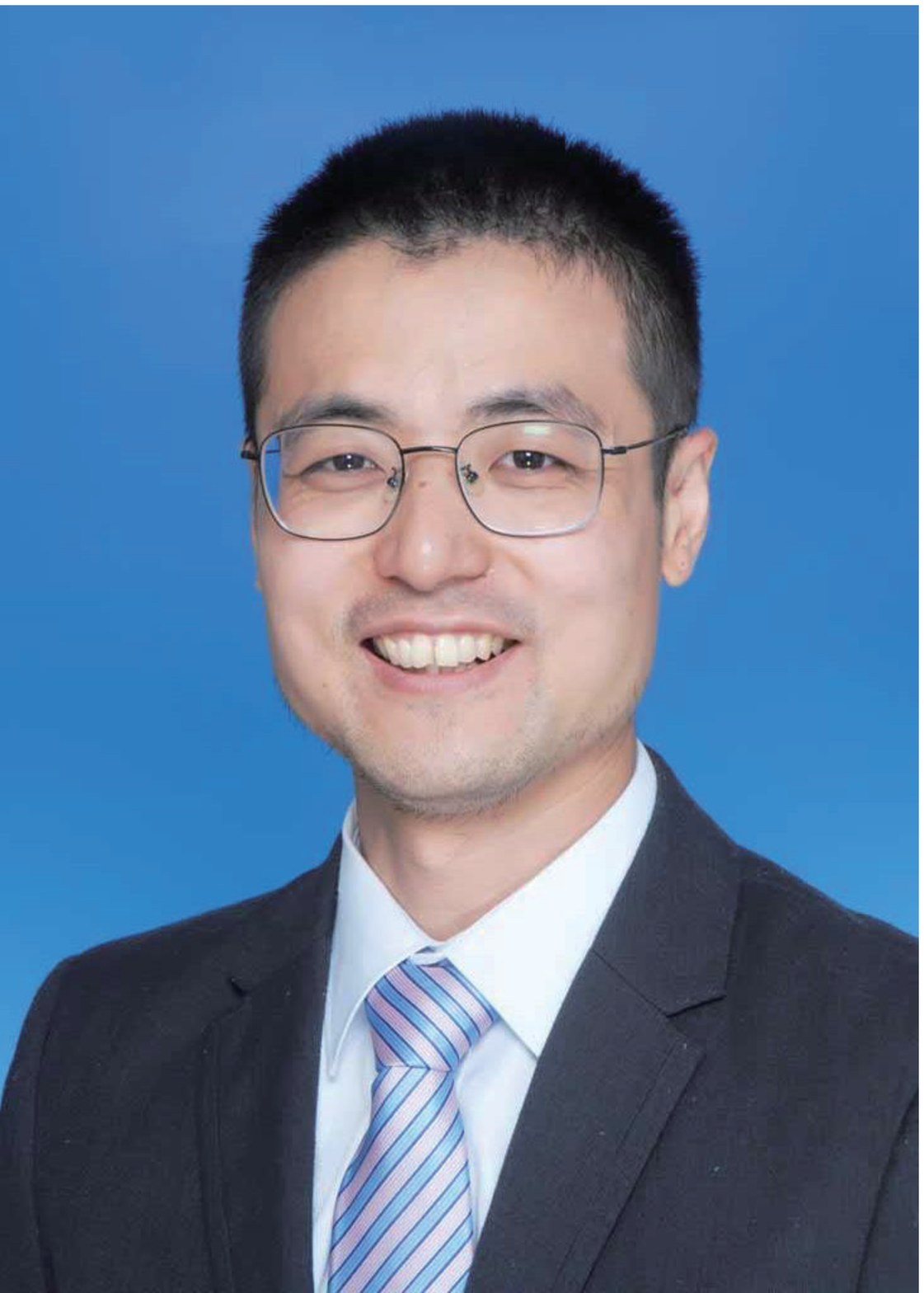}}]
    {Jiayi Zhang}(SM'20) received the Ph.D. degree of Communication Engineering from Beijing Jiaotong University, China in 2014. Since 2016, he has been a Professor with School of Electronic and Information Engineering, Beijing Jiaotong University, China. From 2014 to 2016, he was a Postdoctoral Research Associate with the Department of Electronic Engineering, Tsinghua University, China. From 2014 to 2015, he was also a Humboldt Research Fellow in Institute for Digital Communications, Friedrich-Alexander-University Erlangen-N\"urnberg (FAU), Germany. His current research interests include cell-free massive MIMO, reconfigurable intelligent surface (RIS), communication theory and applied mathematics.

    Dr. Zhang received the Best Paper Awards at the WCSP 2017 and APCC 2017, the URSI Young Scientist Award in 2020, and the IEEE ComSoc Asia-Pacific Outstanding Young Researcher Award in 2020. He was recognized as an exemplary reviewer of the \textsc{IEEE Communications Letters} in 2015-2017. He was also recognized as an exemplary reviewer of the \textsc{IEEE Transactions on Communications} in 2017-2019. He was the Lead Guest Editor of the special issue on ``Multiple Antenna Technologies for Beyond 5G" of the \textsc{IEEE Journal on Selected Areas in Communications}. He was the Editor of \textsc{IEEE Communications Letters} from 2017-2021. He currently serves as an Associate Editor for \textsc{IEEE Transactions on Communications}.
\end{IEEEbiography}

\begin{IEEEbiography}[{\includegraphics[width=1in,height=1.25in,clip,keepaspectratio]{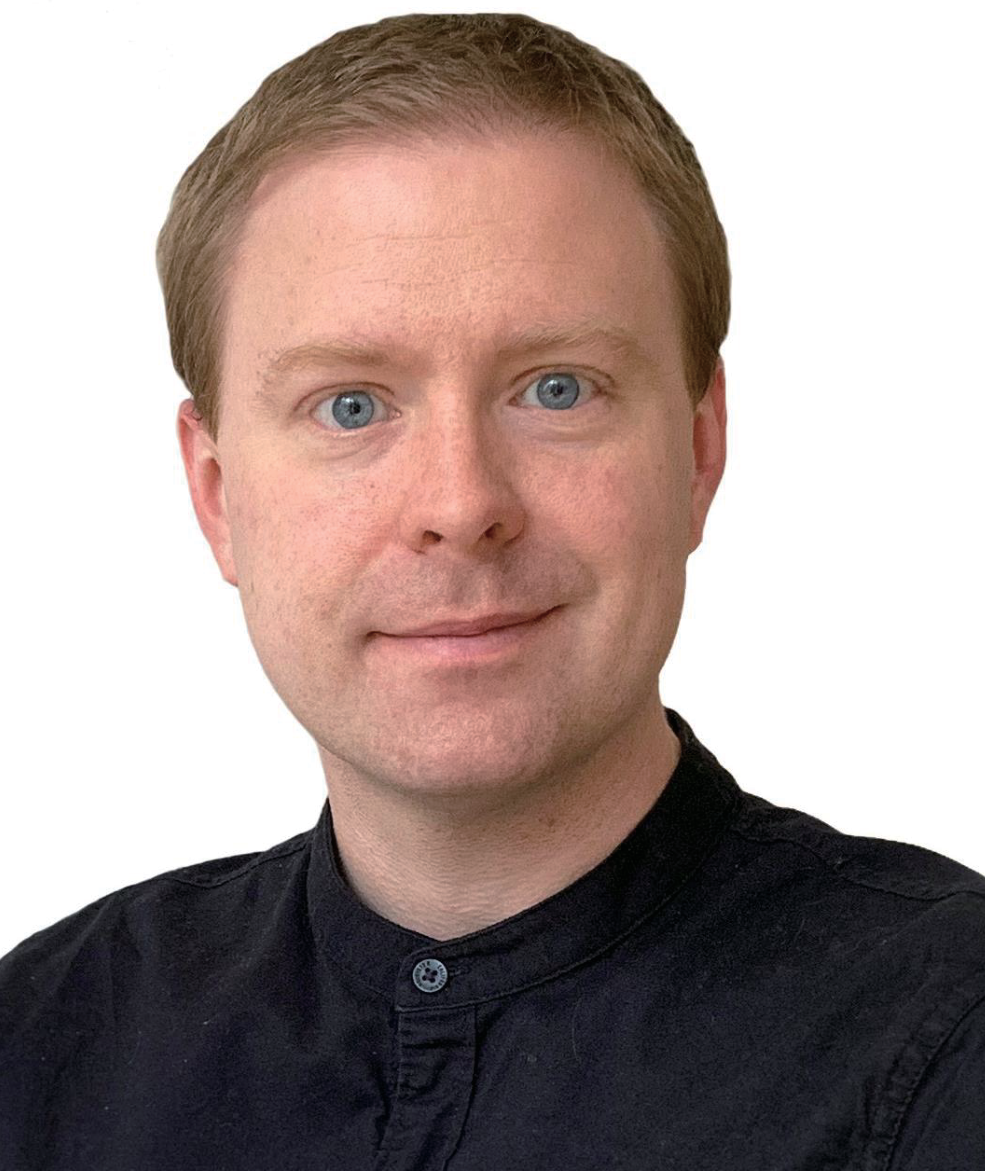}}]
    {Emil Bj{\"o}rnson} (S'07-M'12-SM'17-F'22) is a Full Professor of Wireless Communication at the KTH Royal Institute of Technology, Sweden. He received an M.S. degree in engineering mathematics from Lund University, Sweden, in 2007, and a Ph.D. degree in telecommunications from KTH in 2011. From 2012 to 2014, he was a post-doc at the Alcatel-Lucent Chair on Flexible Radio, SUPELEC, France. From 2014 to 2021, he held different professor positions at Link\"oping University, Sweden. He was a Visiting Full Professor at KTH in 2020-2021, before obtaining a tenured position in 2022.

    He has authored the textbooks \emph{Optimal Resource Allocation in Coordinated Multi-Cell Systems} (2013), \emph{Massive MIMO Networks: Spectral, Energy, and Hardware Efficiency} (2017), and \emph{Foundations of User-Centric Cell-Free Massive MIMO} (2021). He is dedicated to reproducible research and has made a large amount of simulation code publicly available. He performs research on MIMO communications, radio resource allocation, machine learning for communications, and energy efficiency. He is an Area Editor in IEEE Signal Processing Magazine.

    He has performed MIMO research for 16 years, his papers have received more than 23000 citations, and he has filed more than twenty patent applications. He is a host of the podcast Wireless Future and has a popular YouTube channel with the same name. He is an IEEE Fellow, a Wallenberg Academy Fellow, a Digital Futures Fellow, and an SSF Future Research Leader. He has received the 2014 Outstanding Young Researcher Award from IEEE ComSoc EMEA, the 2015 Ingvar Carlsson Award, the 2016 Best Ph.D. Award from EURASIP, the 2018 and 2022 IEEE Marconi Prize Paper Awards in Wireless Communications, the 2019 EURASIP Early Career Award, the 2019 IEEE Communications Society Fred W. Ellersick Prize, the 2019 IEEE Signal Processing Magazine Best Column Award, the 2020 Pierre-Simon Laplace Early Career Technical Achievement Award, the 2020 CTTC Early Achievement Award, the 2021 IEEE ComSoc RCC Early Achievement Award, and the 2023 IEEE ComSoc Outstanding Paper Award. He also co-authored papers that received Best Paper Awards at the conferences, including WCSP 2009, the IEEE CAMSAP 2011, the IEEE SAM 2014, the IEEE WCNC 2014, the IEEE ICC 2015, and WCSP 2017.
\end{IEEEbiography}

\begin{IEEEbiography}[{\includegraphics[width=1in,height=1.25in,clip,keepaspectratio]{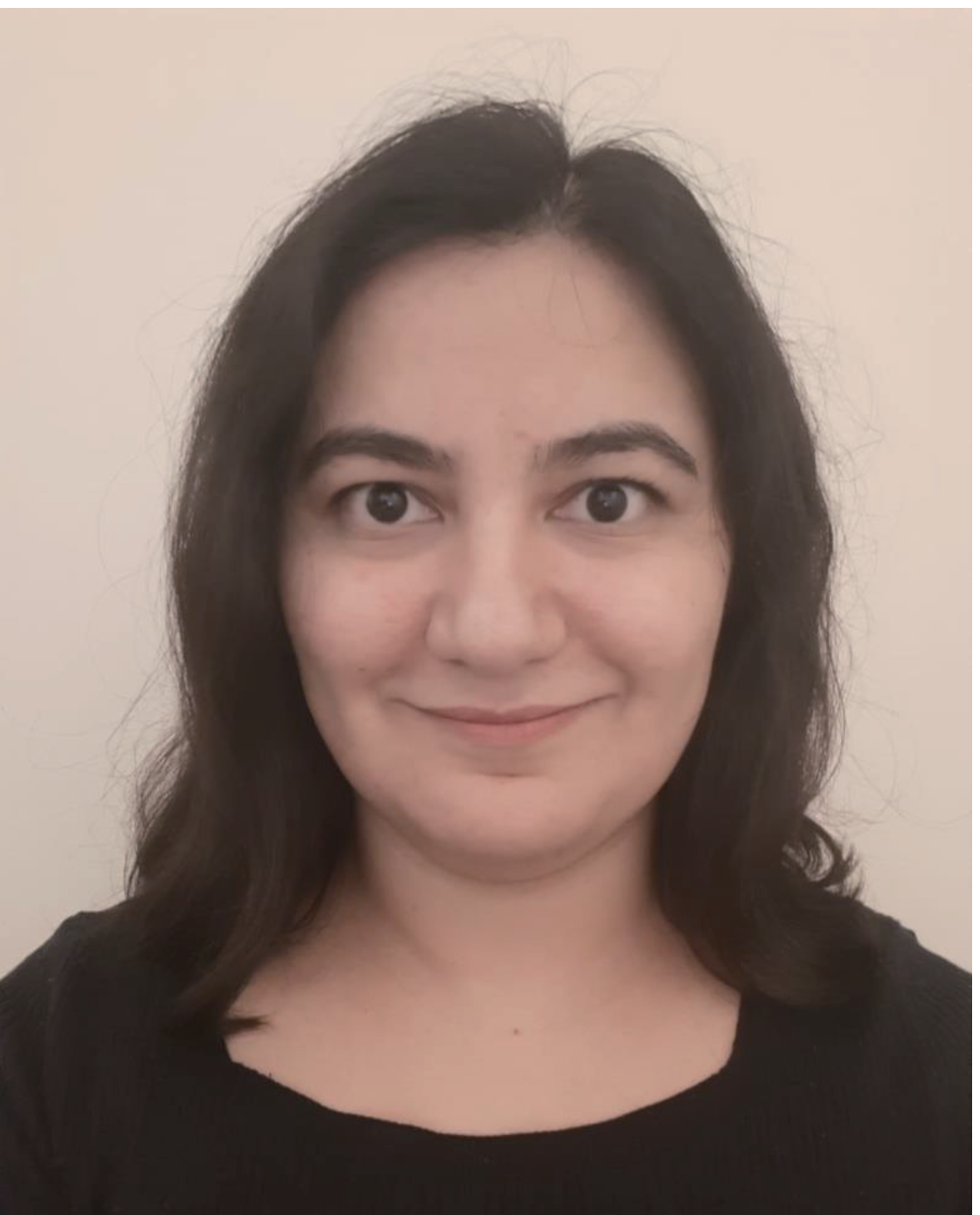}}]
    {{\"O}zlem Tu{\u{g}}fe Demir} received the B.S., M.S., and Ph.D. degrees in Electrical and Electronics Engineering from Middle East Technical University, Ankara, Turkey, in 2012, 2014, and 2018, respectively. She was a Postdoctoral Researcher at Link\"oping University, Sweden in 2019-2020 and at KTH Royal Institute of Technology, Sweden in 2021-2022. She is currently an Assistant Professor with the Department of Electrical and Electronics Engineering, TOBB University of Economics and Technology, Ankara, Turkey. She has authored the textbook \emph{Foundations of User-Centric Cell-Free Massive MIMO} (2021). Her research interests focus on signal processing and optimization in wireless communications, massive MIMO, cell-free massive MIMO, beyond 5G multiple antenna technologies, reconfigurable intelligent surfaces, machine learning for communications, mobile data analysis, and green mobile networks.
\end{IEEEbiography}

\begin{IEEEbiography}[{\includegraphics[width=1in,height=1.25in,clip,keepaspectratio]{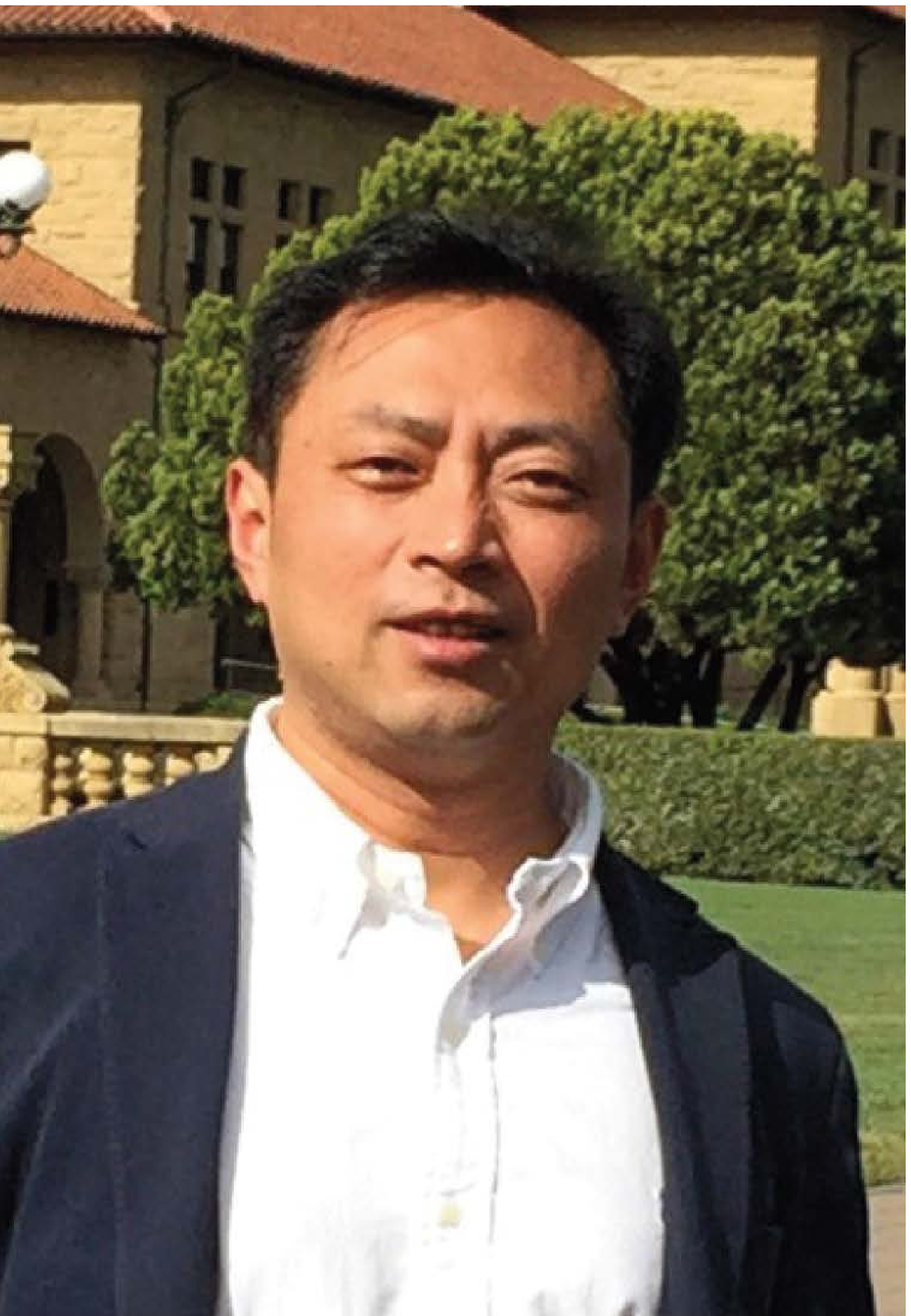}}]
	{Bo Ai}(M'00-SM'10-F'22) received the M.S. and Ph.D. degrees from Xidian University, China. He graduated from Tsinghua University with the honor of the Excellent Postdoctoral Research Fellow in 2007.

    He was a Visiting Professor with the Electrical Engineering Department, Stanford University, Stanford, CA, USA, in 2015. He is currently a Full Professor with Beijing Jiaotong University, where he is the Dean of the School of Electronic and Information Engineering, Deputy Director of the State Key Laboratory of Rail Traffic Control and Safety and the Deputy Director of the International Joint Research Center. He is one of the directors for Beijing Urban Rail Operation Control System International Science and Technology Cooperation Base, and the Backbone Member of the Innovative Engineering based jointly granted by the Chinese Ministry of Education and the State Administration of Foreign Experts Affairs. He is the research team leader of 26 national projects. He holds 26 invention patents. His research interests include the research and applications of channel measurement and channel modeling and dedicated mobile communications for rail traffic systems. He has authored or co-authored eight books and authored over 300 academic research articles in his research area. Five papers have been the ESI highly cited paper. He has won some important scientific research prizes. He has been notified by the Council of Canadian Academies that based on the Scopus database, he has been listed as one of the top 1\% authors in his field all over the world. He has also been feature interviewed by the IET Electronics Letters.

    Dr. Ai is a fellow of The Institute of Electrical and Electronics Engineers (IEEE), The Institution of Engineering and Technology (IET), and an IEEE VTS Distinguished Lecturer. He received the Distinguished Youth Foundation and Excellent Youth Foundation from the National Natural Science Foundation of China, the Qiushi Outstanding Youth Award by the Hong Kong Qiushi Foundation, the New Century Talents by the Chinese Ministry of Education, the Zhan Tianyou Railway Science and Technology Award by the Chinese Ministry of Railways, and the Science and Technology New Star by the Beijing Municipal Science and Technology Commission. He is an IEEE VTS Beijing Chapter Vice Chair and an IEEE BTS Xi’an Chapter Chair. He was a co-chair or a session/track chair of many international conferences. He is an Associate Editor of the \textsc{IEEE Antennas and Wireless Propagation Letters} and the \textsc{IEEE Transactions on Consumer Electronics}, and an Editorial Committee Member of the \textsc{Wireless Personal Communications Journal}. He is the Lead Guest Editor of Special Issues on the \textsc{IEEE Transactions on Vehicular Technology}, the \textsc{IEEE Antennas and Propagations Letters}, and the \textsc{International Journal on Antennas and Propagations}.
\end{IEEEbiography}	

\end{document}